\documentclass[12pt, preprint]{aastex}

\shorttitle{NGC 3603}
\shortauthors{Melena et al}
\slugcomment{Accepted by the Astronomical Journal}
\begin{document}

\title{The Massive Star Content of NGC 3603\altaffilmark{1}}
\author{
Nicholas W. Melena\altaffilmark{2},
Philip Massey\altaffilmark{2},
Nidia I. Morrell\altaffilmark{3},
Amanda M. Zangari\altaffilmark{2,4}
}
\altaffiltext{1}{This paper is based on data gathered with the 6.5 meter Magellan telescopes located at Las Campanas Observatory, Chile.}
\altaffiltext{2}{Lowell Observatory, 1400 W Mars Hill Road, Flagstaff,
AZ 86001; nmelena@lowell.edu, Phil.Massey@lowell.edu}
\altaffiltext{3}{Las Campanas Observatory, The Carnegie Observatories, Colina El Pino s/n, Casilla 601, La Serena, Chile; nmorrell@lco.cl}
\altaffiltext{4}{Research Experiences for Undergraduates (REU) participant, Summer 2007; current address: Wellesley College, 106 Central Street, Wellesley, MA 02481; azangari@alum.wellesley.edu}

\begin{abstract}

We investigate the massive star content of NGC~3603, the closest known giant H~II region.
We have obtained spectra of 26 stars in the central
cluster using the Baade 6.5-m telescope (Magellan I).
Of these 26 stars, 16 had no previous spectroscopy.  We also obtained photometry
of all of the stars with previous or new spectroscopy, primarily using 
archival {\it HST}
ACS/HRC images.  The total number of stars that have been spectroscopically
classified in NGC 3603 now stands at 38.  The
sample is dominated by very early O type stars (O3); there are also several
(previously identified) H-rich WN+abs stars.
 We derive
$E(B-V)=1.39$, 
and find that there is very little variation in reddening across
the cluster core, in agreement with previous studies.  Our spectroscopic parallax
is consistent with the kinematic distance  only if the ratio of total to selective extinction
 is anomalously high within the cluster, as argued by Pandey et al.  
 Adopting their reddening,
we derive a distance of 7.6~kpc.  We discuss the various distance estimates to the
cluster, and note that although there has been a wide range of values in the
recent literature (6.3-10.1~kpc) there is actually good agreement with
the apparent distance modulus of the cluster---the disagreement has been
the result of the uncertain reddening correction. 
 We construct our H-R diagram using the apparent
distance modulus with a correction for the slight difference in differential reddening
from star to star.
The resulting H-R diagram reveals
that the most massive stars are highly coeval, with an age of 1-2~Myr, and of very
high masses (120 $M_\odot$).  The three stars with Wolf-Rayet features are the most
luminous and massive, and are coeval with the non-WRs,  in accord with what was found in  the R136 cluster.
There may be a larger age spread (1-4~Myr)
for the lower mass objects (20-40$M_\odot$).  Two supergiants (an OC9.7 I and the
B1 I star Sher 25) both have an age of about 4~Myr.  We compare the stellar content of this cluster
to that of R136, finding that the number of very high luminosity ($M_{\rm bol}\leq -10$)
stars  is only about 1.1-2.4$\times$  smaller in NGC 3603.  The most luminous members in both
clusters are H-rich WN+abs stars, basically ``Of stars on steroids", relatively unevolved stars
whose high luminosities results in high mass loss rates, and hence spectra that mimic that of
evolved WNs.
To derive an initial mass function for the massive
stars in NGC 3603  requires considerably more spectroscopy; we estimate from a color-magnitude
diagram that less than a third of the stars with masses above 20$M_\odot$ have spectral types known.
\end{abstract}


\section{Introduction}

NGC 3603 is our local giant H~II region, and our window into the giant H~II regions seen in other galaxies. 
Its high luminosity was first recognized by Goss \& Radhakrishnan (1969).
NGC 3603 is  located approximately 7-8 kpc  from the the Sun in the Carina spiral arm.
Eisenhauer et al.\ (1998) estimate that the NGC 3603 cluster has a visual luminosity of $6.1 \times 10^5 L_\odot$,
about a factor of five lower than that of the R136 cluster in the heart of the giant H~II region 30 Doradus in the LMC.
The half-light radii are comparable, 4-5~pc (Eisenhauer et al.\ 1998, and references therein). 
Its relative proximity allows us a unique opportunity to 
study the dense stellar cores that aren't possible in more distant, unresolved regions.  Its stellar content also serves
as a interesting comparison to what is known in R136 (e.g., Hunter et al.\ 1996;
Massey \& Hunter 1998). 
 
To characterize the massive star population of this high luminosity H~II region requires both photometry
and spectroscopy. 
The first photometric study of NGC 3603 was by  Sher  (1965), who 
presented {\it UBV} photometry
(both photoelectric and photographic)  for many of the stars in the periphery of 
the cluster. Sher's (1965) survey was followed by van den Bergh (1978)'s
photoelectric study, and by
Melnick et al.\ (1989), who used {\it UBV} CCD images
to derive photometry for many of the Sher (1965) stars as well as for some stars
located closer to to the clusters' core. Melnick et al.\ (1989) derived a distance of 7.2~kpc, 
and an age of 2-3~Myr but with evidence of considerable age spread.
Moffat et al.\ (1994) were able to use {\it HST} to obtain F439W (essentially {\it B}-band)
photometry of stars even closer to the core, 
although the images were obtained with the original WF/PC camera
and suffered from the effects of the well-known spherical aberration. 
Near-IR photometry
obtained with adaptive optics is discussed by Eisenhauer et al.\ (1998). 
Deep JHKL photometry extending down to the pre-main-sequence population has
been obtained by Stolte et al.\ (2004), and deep optical photometry has been
discussed by Sagar et al.\ (2001) and Sung \& Bessell (2004).   Harayama et al.\ (2007)
further explore the low-mass IMF.   Pandey et al.\ (2000) used multi-color photometry to
explore the reddening within the cluster, finding that it was anomalously high.

Because of crowding, photometry has been hard enough, but spectroscopy has been all but
impossible, especially in the central core.  HD 97950, the core of NGC 3603, has been
known to be composite for many years, similar to the situation for R136a in 30 Doradus
(see Moffat 1983).  It has a collective spectral type of around WN6 + 05 (Walborn 1973). 
Moffat (1983) classified 13 objects in the cluster with photographic spectra, 
finding 11 O-type stars,
one early B supergiant, plus the unresolved WR core.  Some of these spectra
were of blends that were not resolved.   
 With the Faint Object Spectrograph (FOS) on {\it HST}, Drissen
et al.\ (1995) were able to classify 14 objects, 11 of which were early
O-type stars, including five of the earliest class at that time, O3.
The spectra of four stars have been modeled: Crowther \& Dessart (1998) have
analyzed spectra of the three Wolf-Rayet stars in the core, and Smartt et al.\ (2002)
have studied the B supergiant Sher 25.

Here matters have stood until the present.   Two of the authors 
(PM and NIM) proposed unsuccessfully several times
to obtain spectra of stars in the crowded core of
NGC 3603 with the Space
Telescope Imaging Spectrograph (STIS) on {\it HST}, and after STIS's demise it
appeared to many of us that the secrets of NGC~3603 would be well-kept for many
years to come.  However,  modern ground-based telescopes now are capable of
sub-arcsecond image quality, allowing work from the ground that at one time was
only possible from space.
 Here we utilize the good image quality and large aperture of the 
 Baade 6.5-m telescope for spectroscopy of individual stars inwards to the dense 
 central core of NGC 3603, classifying 26 stars, bringing the total number of stars
 with classification to 38.  In addition, we supplement these spectroscopic data with
 photometry from recently obtained {\it HST} images.
These data
permit a better characterization of the massive star population of this nearby giant H~II 
region than
has hitherto been possible.  
In \S~\ref{Sec-obs} we will describe the observations and reductions. In \S~\ref{Sec-types}
we describe our spectral classifications.  In \S~\ref{Sec-results} we will
use the resulting spectral types and photometry to derive new values for the reddening and
distances to the cluster, and construct an H-R diagram.  In \S~\ref{Sec-summary} we
will summarize
and discuss our results.

\section{Observations and Reductions}
\label{Sec-obs}

\subsection{The Sample}

Previous studies have
found that the reddening across NGC~3603 is large
($E(B-V)\sim 1.4$) but quite uniform (Moffat 1983), so that
the visually brightest stars are likely to have the highest visual absolute magnitudes.
Since the cluster is dominated by early O-type stars, optical photometry cannot tell
us much about the physical properties of the stars, as the colors are degenerate
with effective temperature (see Massey 1998a, 1998b), but does allow one to weed
out 
any obvious non-members  as was done by
Melnick et al.\ (1989).  Thus we decided to obtain
photometry and spectroscopy for as many of the visually brightest members as was
practical.  To achieve this goal, we identified our sample using {\it HST} images that had been
obtained with
the High Resolution Camera (HRC) of the Advanced Camera for Surveys (ACS)
in connection with a different study of NGC 
3603\footnote{ Obtained under program ID 10602, PI=Jes\'us Ma\'{\i}z Apell\'aniz.}.
We also use these images for our photometry (\S~\ref{Sec-phot}).

In Table~\ref{tab:sample} we list the sample of stars for which we have 
obtained either spectroscopy or photometry.  We were able to obtain photometry
for all of the stars with  spectroscopy (both our own spectroscopy and previous), and
so Table~\ref{tab:sample} contains
all of the NGC~3603 stars with optical spectroscopy to the present\footnote{This excludes the Moffat
(1983) spectra of blobs given letter designations.}.
Of these, 26 were obtained by ourselves in the present study.  We have included a few
stars with new photometry for which we lack spectroscopy as yet.  The coordinates in
Table~\ref{tab:sample} have been determined from
the ACS images and our own ground-based images, where care has been taken to
place these on the UCAC2 system (Zacharias et al.\ 2004).
 We have retained previous nomenclature for stars
with previously published spectroscopy or for stars clearly identified in Sher (1965), 
but difficulties in identification from the finding charts published by 
Melnick et al.\ (1989) and Moffat et al.\ (1994) have resulted in our presenting new
designation for other stars.  All stars in our sample are identified in Figs.~\ref{fig:fc1}
and \ref{fig:fc2}.

\subsection{Spectroscopy}

Optical spectroscopy was obtained by NIM during two nights with the Baade 6.5-m Magellan
telescope on Las Campanas: 2006 April 12 and 15.  The data were taken with the
Inamori Magellan Areal Camera and Spectrograph (IMACS) in its long camera mode
(f/4) using the 600 line mm$^{-1}$
grating.  The wavelength coverage was approximately 3600\AA\ to 6700\AA, with
a dispersion of 0.37\AA\ pixel$^{-1}$, and
a spectral resolution of 2.0\AA. 
The detector for IMACS consists of a mosaic of 8 CCDs.  In our case the spectra
fell only along one row of 4 CCDs, and the camera was oriented such that
the dispersion crossed all four CCDs, leaving three narrow (20\AA)
gaps in our coverage: 4330-4350\AA\ (which included H$\gamma$), 5120-5140\AA,
and 5920-5940\AA.  A long 0\farcs7 slit was used.  

The spectrograph was 
usually rotated so that we would obtain two or more stars of interest on the slit at
the same time.  A direct image was usually obtained immediately before the
spectroscopic exposure, allowing us to be quite certain of the identification of stars
for which we obtained data, including ones that were coincidentally observed by their
falling by chance on the slit.  
Flat field exposures were obtained at the beginning and/or end of each night, and
HeNeAr comparison arc exposures were taken for each new position.  The 
spectroscopic exposures typically consisted of three individual exposures in order
to facilitate the removal of cosmic rays.  Exposure times ranged from 3$\times$250 s
to 3$\times$600 s.

The seeing was good but not spectacular.  Using the direct images we 
obtained adjancent to the spectroscopic exposures, we measured an average
full width at half maximum of 0\farcs9 on the first night, and 0\farcs8 on
the second.  The best images were 0\farcs67, and the worst were 1\farcs1.

Each chip and exposure were reduced separately, and the data were
combined at the end to produce a single spectrum for each object along the
slit.   The processing steps were the usual ones and were done using IRAF\footnote{IRAF is 
distributed by the National Optical Astronomy Observatory, which
is operated by the Association of Universities for Research in Astronomy, Inc., under
cooperative agreement with the National Science Foundation (NSF). We are grateful to the
on-going support of IRAF and the help ``desk" maintained by the volunteers at
http://www.iraf.net.}.
First, the overscan was used to remove the bias on each chip, and then the 
the two-dimensional bias structure was subtracted using the average of 10 
zero-second exposures.  The flat field exposures were divided into the data, after normalization.
The spectra were extracted using an optimal extraction algorithm, after defining the 
location of stars of interest and selecting the sky background regions interactively. 
The same trace (spatial position as a function of wavelength) and extraction apertures were
then applied to the comparison arcs, and a wavelength solution was found for each aperture.
The stellar spectra were then normalized to the continua, combined, and the four wavelength
regions merged into a single spectrum for each star. 

The typical signal to noise ratio (SNR) is 200-500 per 2\AA\ spectral resolution element.  Such
good quality spectra are essential for detecting very weak He~I in the earliest O-type
stars.

\subsection{Photometry and Transformation Issues}
\label{Sec-phot}

Most of our stars were present in the {\it HST} ACS/HRC  field of view (Fig~\ref{fig:fc2})
 and this is our primary source
for the photometry.  The images we analyzed  had been taken through the F435W and F550M filters,
similar to Johnson $B$ and $V$ (but see below).  Each image consisted of four dithered exposures
(offset one from another by a fraction of an arcescond),
combined into a single ``drizzled" image for better sampling of the point-spread function.  The
scale of the final image is 0\farcs025 pixel$^{-1}$.  The total integration time for each image was 8 s.
The images were taken on 2005 December 29.

For the actual photometry we used the PHOT application of the DAOPHOT package of IRAF, adopting
an aperture radius of 3.0 pixels (0\farcs08).  We then determined the aperture correction from 3.0
pixels to the ACS ``standard" aperture of 0\farcs5 using a few isolated stars on each frame.
We adopted the published aperture correction from 0\farcs5 to infinity
from Sirianni et al.\ (2005).   The counts were corrected for the charge transfer efficiency losses, dependent 
upon the sky background and position on the chip, following the formulation in Pavlovsky et al.\ (2006).
We note that the CTE correction is quite significant, amounting to $-0.01$ to $-0.07$~mag, depending
upon the brightness and location of a star.

Although Sirianni et al.\ (2005)
recommend working in the ACS ``native" photometric system, we have
instead chosen to transform to the standard (Johnson)
system.
We wish to determine the color excesses and distance modulus
to the cluster, and to do this we rely upon our knowlege of
the intrinsic colors and absolute magnitudes of
O stars as a function of spectral type (see, for example,
Conti 1988 and Massey 1998a). These quantities
are known {\it empirically} as the result of observational studies
in the standard system rather than through models that
can simply be recomputed for some alternative bandpass.  In other
words we  know that an O7~V star has $(B-V)_0$ of $-0.32$ and
$M_V=-4.9$ through observations of young clusters, not as the
result of computations.
Thus, in order to compute color excesses and derive
spectroscopic parallaxes we need to transform {\it something}:
either we transform the intrinsic colors and absolute magnitudes to the
ACS system, or we transform the ACS photometry to the standard system.  We
have chosen the latter, since (a) this allows us to include stars outside
the ACS/HRC field for which standard photometry exists, and (b) it allows
us to compare our results to others.  That said, we need to emphasize that the
transformations introduce uncertainities well beyond the usual photometric
errors.

Sirianni et al.\ (2005) list transformations to {\it B} for F435W ACS/HRC photometry,
but the transformations are based upon unreddened stars, rather than reddened
O stars as is the case here.  Furthermore, they offer no transformation
for the F550M filter.  So, we have determined corrections ourselves
as follows.  First,
we placed the F435W and F550M photometry on the so-called ``VEGAMAG"
system using the revisions to the Sirianni et al.\ (2005) zeropoints listed on
the ACS web site\footnote{http://www.stsci.edu/hst/acs/analysis/zeropoints.}.
We then corrected the photometry for the fact that Vega actually doesn't
have zero
magnitude, but rather $B=V=+0.03$ (Bessell et al.\ 1998; see also
Ma\'{\i}z Apell\'aniz 2006).
Using the Kurucz (1992) Atlas 9 models
appropriate for
O-type
stars ($\log g=5.0$ and $T_{\rm eff}=$40,000, 45,000, and 50,000 K) we
then computed $F550M - V$ and $(F435W-F550M) -(B-V)$ for various amounts
of reddenings,
where we have adopted the Cardelli et al.\ (1989) Galactic reddening law.

For an O star with reddening in the range we expect for NGC 3603 [i.e.,
$E(B-V)=1.3-1.5$],
we find $F550M-V=-0.11$ to $-0.12$.
We adopt $F550M-V=-0.11$.  Similarly we find $F435W-B=+0.0$ to $+0.02$,
and we adopt $F435W-B=+0.01$. Thus $(F435W-F550M)-(B-V)=+0.12$\footnote{In 
contrast, for an {\it unreddened} O-type star we find
 $F550M-V=+0.0$,  $F435W-B=-0.07$
and hence $(F435W-F550N) - (B-V)=-0.07$.}.
Jes\'{u}s Ma\'{\i}z Apell\'{a}niz (private
communication) derives essentially identical corrections for similarly
reddened O stars.  

We note that the transformations are
sensitive at the 0.02-0.03~mag level to the details of the adopted bandpasses.  For the ACS bandpasses
we synthesized an effective bandpasses using SYNPHOT, which includes
the wavelength-dependent sensitivities of the entire system (telescope + filter + instrument).  
Historically, knowledge of the standard Johnson bandpasses has required some
``reverse engineering", tested by using the deduced bandpasses with model atmospheres
to reproduce the
observed colors of stars (see Buser \& Kurucz 1978 and Bessell et al.\ 1998).  
For $B$ and $V$ SYNPHOT adopts the bandpasses determined
by Ma\'{\i}z Apell\'aniz (2006), which are similar to, but not quite identical, to the
Bessell (1990) versions.  Using the Bessell (1990) prescriptions would result in a 0.02~mag shift
in both the $F435W-B$ and $F550M-V$ transformations used here.   Similarly,
 if we instead adopted
the  bandpasses determined by
Buser \& Kurucz (1978), our conversions would differ by 0.03~mag for
$F435W-B$ and by 0.01 in $F550M-V$.  

We list the resulting magnitudes  and colors
in Table~\ref{tab:sample} as {\it V} and {\it B-V}.
It is worth comparing our photometry to that of
others, as a check on our transformations.  For 9 stars  in common to Melnick et al.\ (1989) in
our $V$-band photometry, we exclude two outliers, and then
find a mean difference of +0.04~mag, in the sense of us minus Melnick et al.\ (1989).
We identify only three stars in common for which we have $B-V$ values; for these, we find an
average difference of $-0.02$~mag in the same sense. 
Drissen et al.\ (1995)
list ``$B$" values they derived from just averaging  the flux in their spectra from 4000\AA\ to 4750\AA.
The difference for their 13 stars (us minus them) is +0.05~mag\footnote{We exclude the star NGC3603-33, which both they and we find to be marginally resolved on the {\it HST} images.}.
We conclude that our photometry is the best that we can do unless and until well-calibrated 
exposures can be made under good seeing conditions in filter systems which are a
closer approximation to the standard system.

\begin{figure}
\plotone{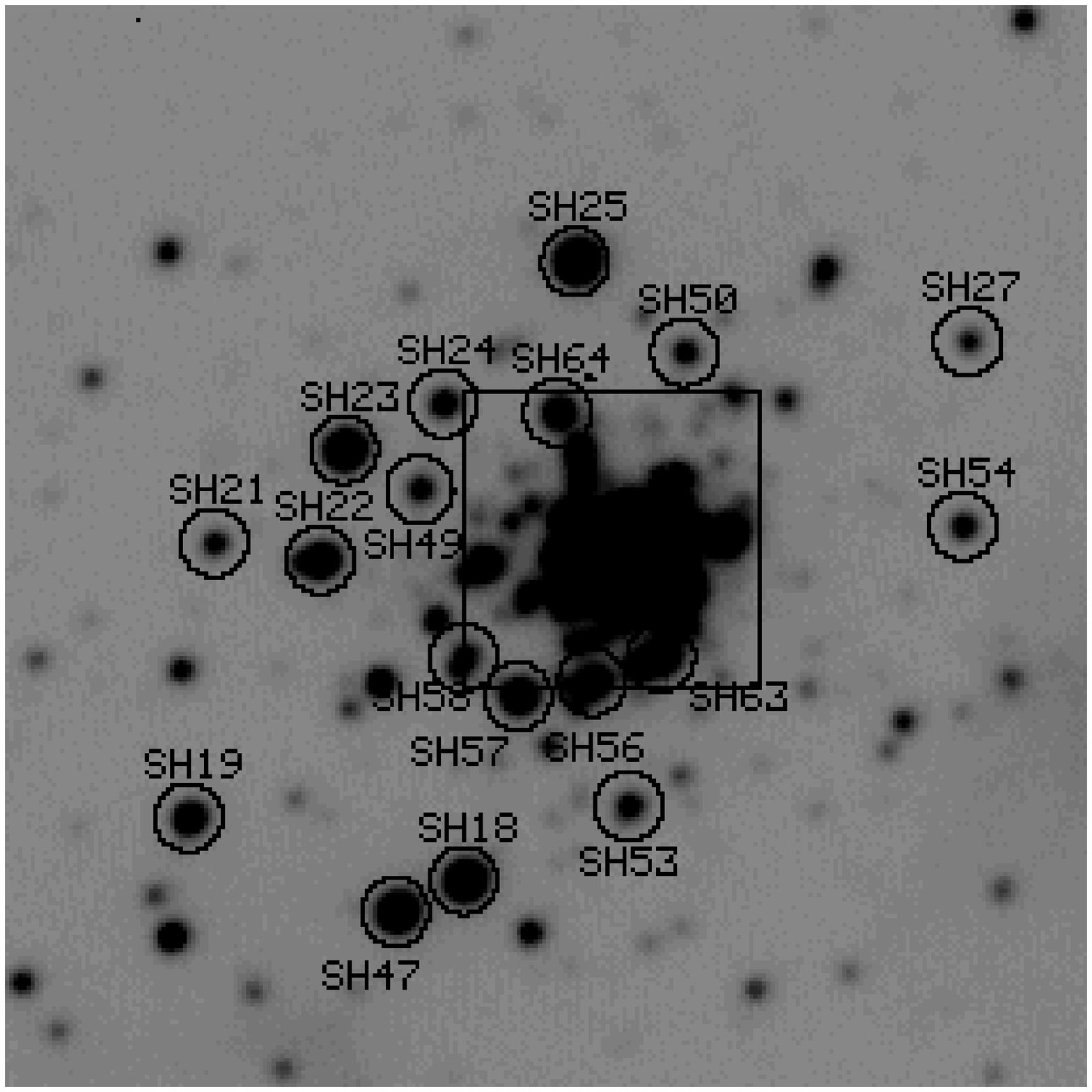}
\caption{\label{fig:fc1}.  Finding chart for stars in our sample from
the periphery of NGC 3603.
The field of view is about 1\farcm0 on a side.  North is up, and east is to the left. The chart
is made from a V-band image taken with the CTIO Yale 1.0-m telescope. The circles
are 4\farcs6 in diameter, or 0.17~pc at an assumed distance of 7.6~kpc. The square denotes the
approximate area shown in Fig.~\ref{fig:fc2}.}
\end{figure}

\begin{figure}
\plotone{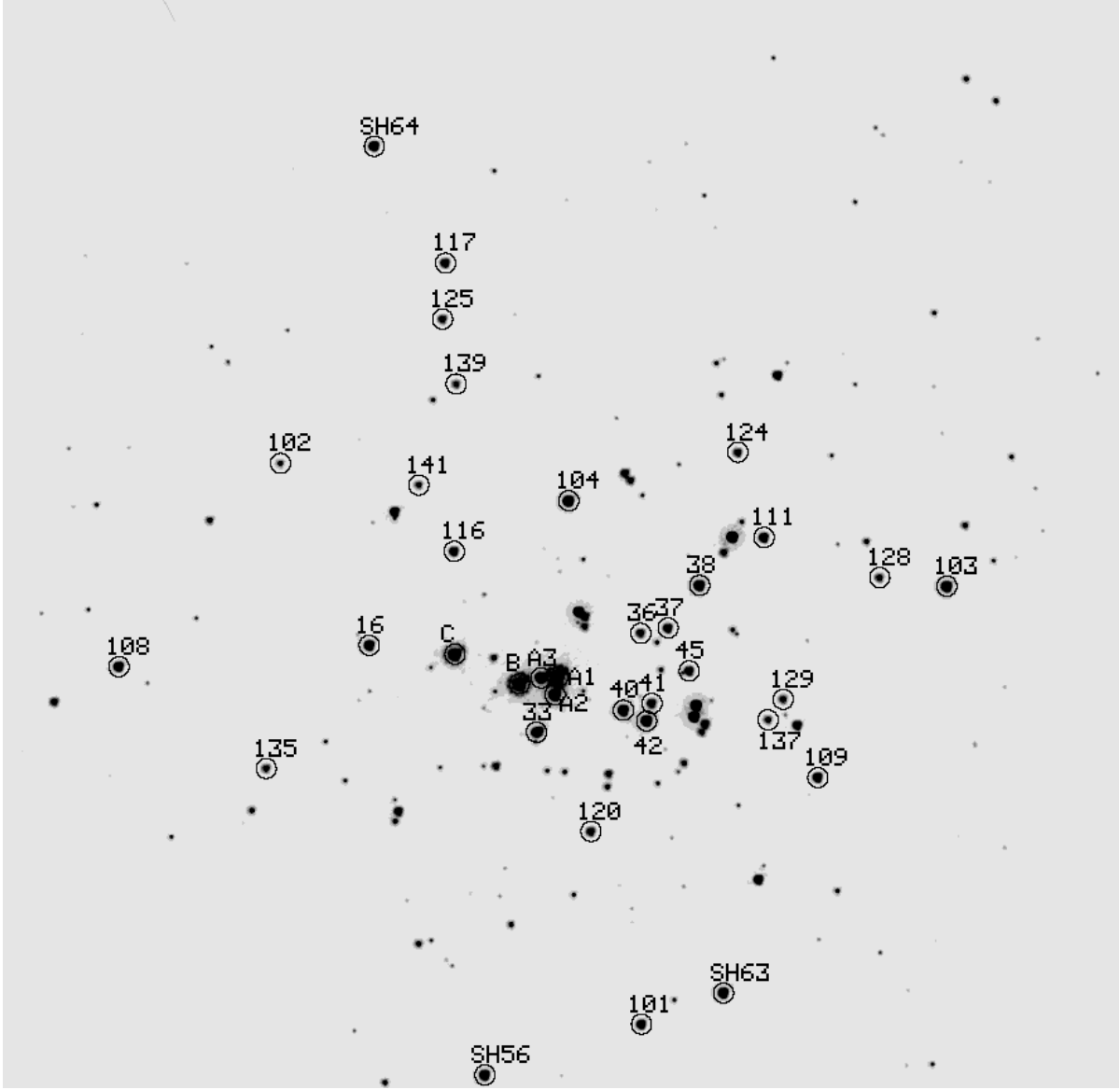}
\caption{\label{fig:fc2}.  Finding chart for stars in our sample from the center
of NGC 3603.  The field of view is 20\arcsec on a side, and roughly corresponds to the
square outline shown in Fig.~\ref{fig:fc1}. The image has been rotated
so that north is up and east is to the left.  The chart was made from a
V-band ACS/HRC image.  The circles are  0\farcs4 in diameter, or 0.02~pc at an assumed distance of 7.6~kpc.}
\end{figure}

 Some stars fell  outside the area covered by the HRC image.  Although the same {\it HST}
 ACS program included images taken with the Wide Field Camera of ACS, all the stars of interest
 were quite saturated on these archival images.  So, instead we used a 20 s V-band exposure
 taken for us by the SMARTS consortium as part of a separate investigation looking for eclipsing
 binaries in NGC 3603.  We used three stars in common to set the photometric zero-point consistently
 with the ACS/HRC system.  The image was taken on 2006 April 3, and the seeing was 1\farcs4, and is
 used for the finding chart in Fig.~\ref{fig:fc1}.
 As we lacked a {\it B}-band exposure, we adopt the $B-V$ values of Melnick et al.\ (1989) for those
 stars, when available.   The agreement between the SMARTS data and Melnick et al.\ (1989) is good,
 with an average difference of +0.03~mag, again in the sense of our 
 values minus theirs.  The photometry
 of the star Sher 58 clearly differs significantly ($\Delta V=0.58$ mag).  Their color of this star leads them to
 conclude it is not a member, but their value is inconsistent with the O8~V spectral type we find, leading us
 to suspect that we have identified different stars as Sher 58.  The star has a close companion, as
 is seen in Fig.~\ref{fig:fc1}.

\section{Spectral Classifications}
\label{Sec-types}

\begin{figure}
\epsscale{0.8}
\plotone{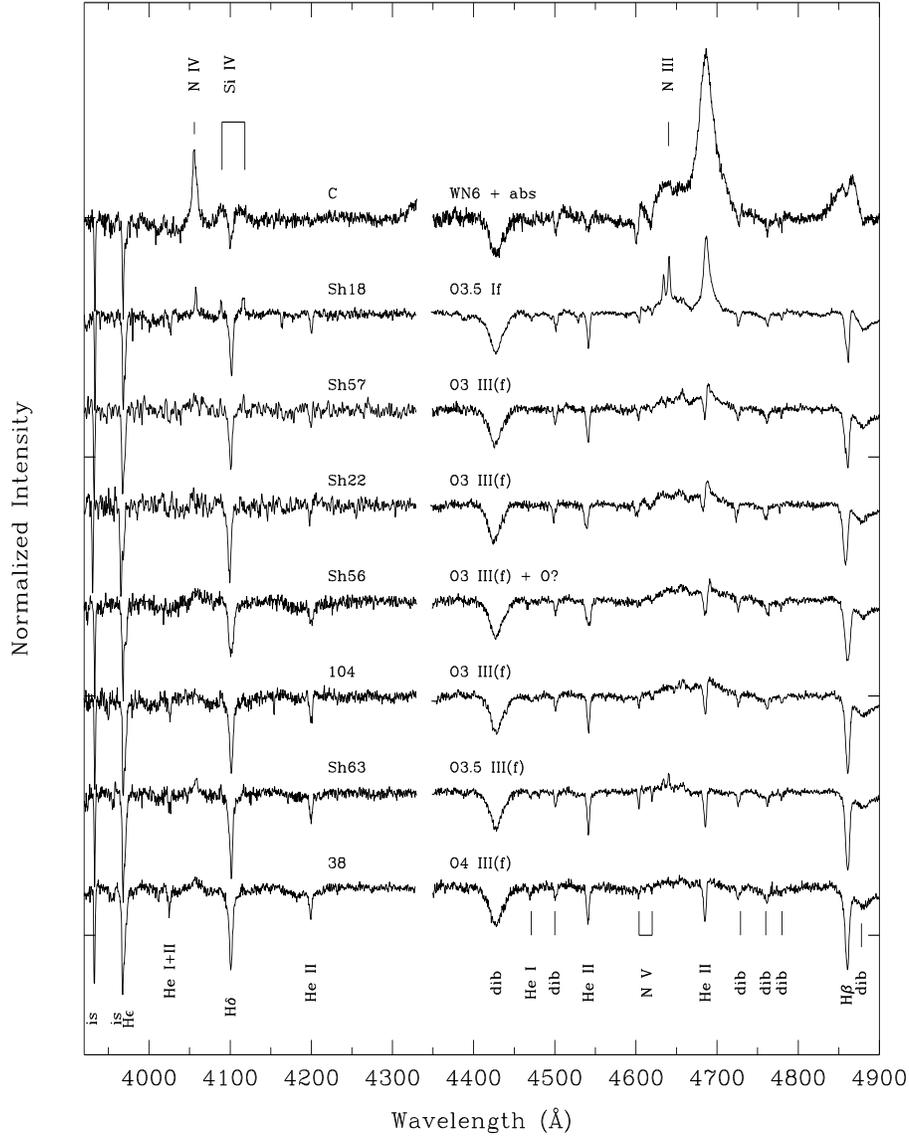}
\caption{\label{fig:spec1}Normalized spectra of NGC 3603 supergiants and giants.
Only the data in the blue (MK classification) region are shown. 
The interstellar (is) H and K Ca II lines and diffuse interstellar bands (dib) are
marked, along with the prominent stellar features.  The displacement between the
various spectra is 0.4 times the continuum level.}
\end{figure}

\begin{figure}
\epsscale{0.8}
\plotone{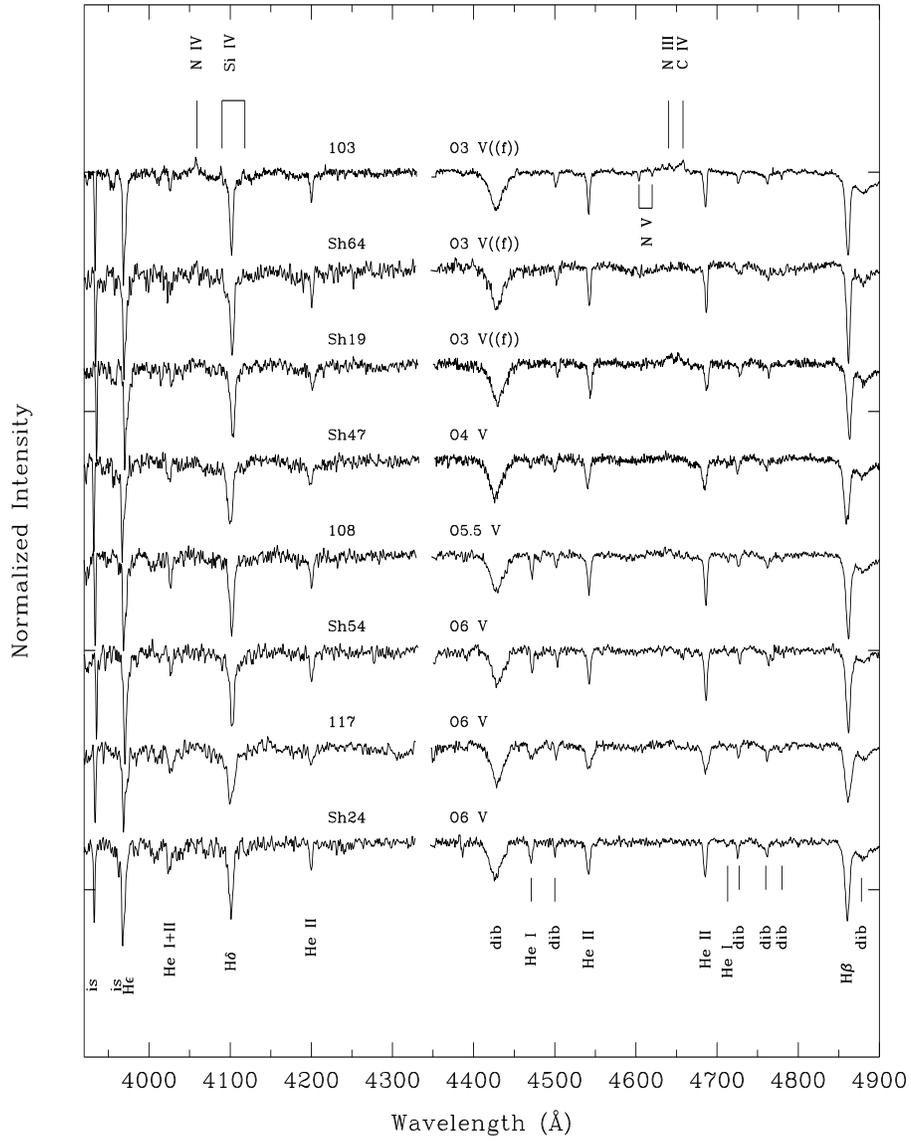}
\caption{\label{fig:spec2} Same as Figure~\ref{fig:spec1} for O3-O6 
dwarfs.}
\end{figure}

\begin{figure}
\epsscale{0.8}
\plotone{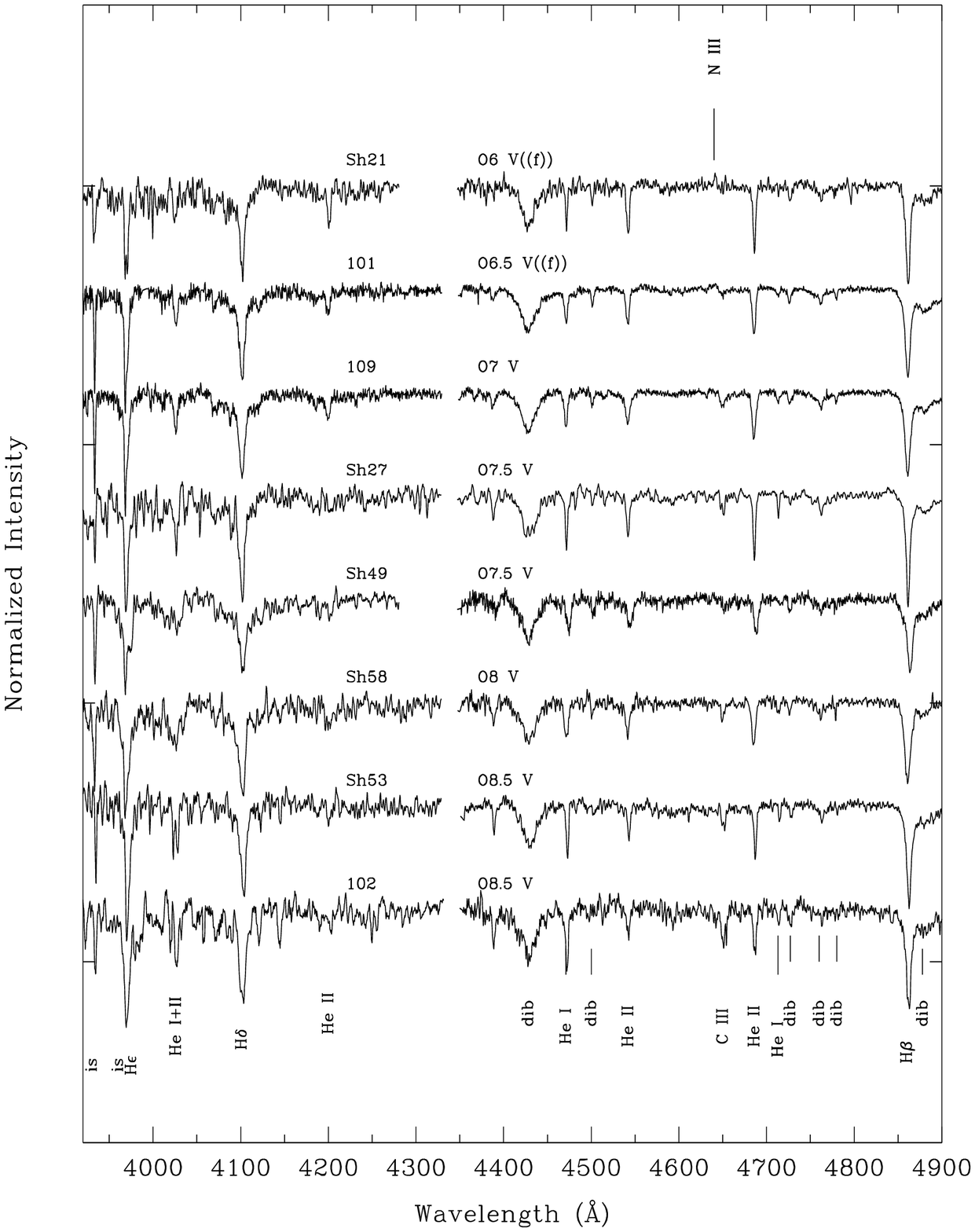}
\caption{\label{fig:spec3} Same as Figure~\ref{fig:spec1} for O6-O8.5 dwarfs, except that the displacement
between the spectra is 0.6 times the continuum level.}
\end{figure}

\begin{figure}
\epsscale{0.8}
\plotone{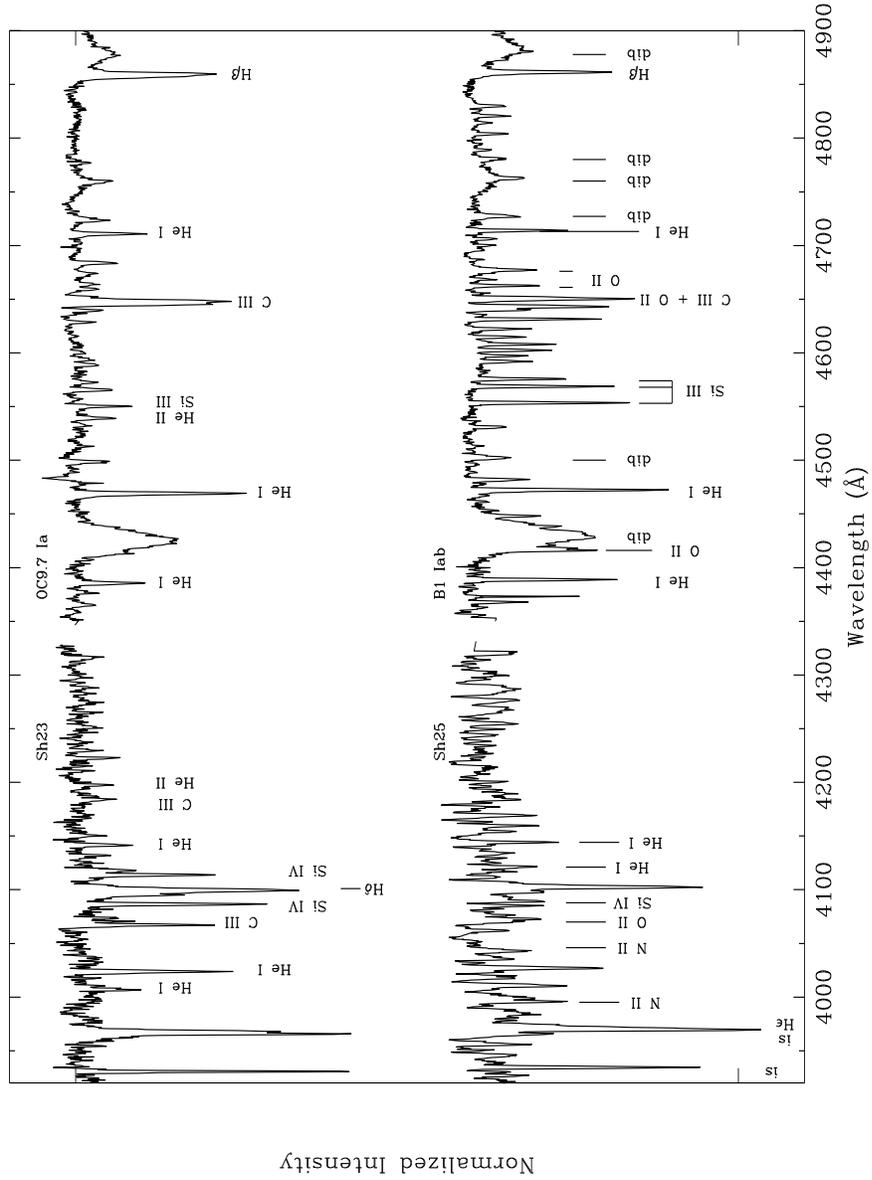}
\caption{\label{fig:spec4} Same as Figure~\ref{fig:spec1} for the two ``late-type" supergiants
in our sample, and OC9.7 Iab star (Sh 23) and a B1 Iab star (Sh 25).  The bluest part of the
spectrum 
in the latter star has been heavily smoothed. The displacement between the two spectra is 0.6 times the
continuum level.}
\end{figure}

The spectra were classified two ways.  First, a qualitative assessment was made
by comparing the observed spectra to those illustrated by Walborn \& Fitzpatrick \ (1990).
Second, a more precise determination was then made quantitatively by
measuring the equivalent widths (EWs) of He~I $\lambda 4471$ and He~II
$\lambda 4542$ and computing $\log W'=\log$(EW He~I/EW He~II).  The latter
has been calibrated against spectral class by Conti (see Conti \& Alschuler 1971,
Conti 1973,  Conti \& Frost 1977; see summary in
Conti 1988), and indeed forms the astrophysical basis for the
spectral classification of O-type stars.  Our experience has shown that the visual
method works best on spectra with a low or modest SNR, while the latter
is more accurate for data with a high SNR and adequate
spectral resolution.
In either scheme the primary diagnostic for the spectral subtypes for O stars is 
the relative strength of He~I and
 He~II, while the primary luminosity indicators are the strength (emission or absorption) 
 of the He~II $\lambda 4686$ line and the presence and strength of the
 N~III $\lambda \lambda 4634, 42$ emission lines.  For the earliest O types (O2-O3.5) we also considered
 the criteria suggested by Walborn et al.\ (2002), namely the relative strengths of
 N~III and N~IV emission, although it has yet to be shown whether or not this defines an
 extension of the temperature sequence (see discussion in
 Massey et al.\ 2005). We include the new spectral types in
 Table~\ref{tab:sample}, and illustrate our spectra in Fig.~\ref{fig:spec1}-\ref{fig:spec4}.

 In a few cases it was clear that our spectral extraction apertures contained some
 light from neighboring objects.  If the contamination was judged severe, we did not
 include the star in our study.  However, there were a few cases where there was some minor
 blending, and we note such cases.  For these,
 the spectral types may not be as good as for the other stars in our sample.
    
We were able to classify 26 stars in the cluster, 16 of which were previously unclassified. 
Among these are a number of newly classified O3 and O4 types.
We compare our new spectral types to those in the literature in Table~\ref{tab:spec}.
In general we find very good agreement, usually within a single type and luminosity class.
Our experience has shown that nebular contamination can substantially alter the classification, particularly for the early O stars
where He~I $\lambda 4387$ is not visible, and one must rely solely on He~I $\lambda 4471$, which may be filled in by nebular emission. 
In choosing which stars to observe we generally decided not to observe
the stars that had been previously observed
with {\it HST's} FOS
by Drissen et al.\ (1995).  Although the Drissen et al.\ (1995) sample were observed
 without the benefits of sky
subtraction, the FOS  observations were obtained
with such a small aperture  (0\farcs25) that nebular contamination should not be an issue.
However, we did decide to include a number of stars that had been
previously observed by Moffat (1983), as these were observed
photographically.  Given this, we find the agreement remarkably good,
a testament to the careful work of the previous studies.
That said, there are a few differences we note in discussing the stars individually.

{\it NGC3603-Sh27}.  Our spectrum of this star (Fig.~\ref{fig:spec3}) is of unusually low signal-to-noise,
about 130 per 2\AA\ spectral resolution element.  Nevertheless, it is easy to classify
as roughly O7~V. We measure
an EW of 650m\AA\ for He~I $\lambda 4471$, and 600m\AA\ for He~II $\lambda 4542$,
yielding a $\log W'=+0.0$ and thus an O7.5 type.  The lack of He~II $\lambda 4686$ emission makes
this a dwarf.

{\it NGC3603-Sh54}. Visually we classified this star as
O6~V, with He~I $\lambda 4471$ just a bit weaker than He~II $\lambda 4542$ (Fig.~\ref{fig:spec2}).
We measure 310m\AA\ and 520m\AA, respectively, leading to a $\log W'=-0.23$,
also leading to an O6 type (Conti 1988).  The strength of He~II $\lambda 4686$ absorption
and lack of discernible N~III $\lambda \lambda 4634, 42$ emission identifies the star as a dwarf.

{\it NGC3603-103}. The spectrum is shown in Fig.~\ref{fig:spec2}.
This star was slightly blended in our extraction window.  We 
detect extremely weak He~I $\lambda 4471$ absorption in our spectrum (EW=30m\AA),
typical of the very earliest O type stars (see Massey et al.\ 2004, 2005).  He~II
$\lambda 4542$ has an EW of 660m\AA, so $\log W'=-1.3$.  We call this star an O3 V((f)), 
where the luminosity designation denotes a dwarf with weak N~III $\lambda \lambda 4634, 42$ emission and He~II $\lambda 4686$ in absorption.

{\it NGC3603-109}.  Visually the spectrum of this star appears to be of an O7~V, with He~I
$\lambda 4471$ a bit weaker than He~II $\lambda 4542$ (Fig.~\ref{fig:spec3}).  We measure
EWs of 650m\AA\ and 780m\AA, respectively, or $\log W'=-0.08$.   This is consistent
with the O7~V designation.  The star was slightly blended on the slit.

{\it NGC3603-Sh63}.  We show the spectrum of this star in Fig.~\ref{fig:spec1}.
This star has weak He~I $\lambda 4471$, with a EW of about 180m\AA,
a little large for us to consider the star an O3 (i.e., Kudritzki 1980, Simon et al.\ 1983; see discussion in
Massey et al.\  2004, 2005).  Yet, N~IV $\lambda 4058$ emission is much stronger
than that of N~III $\lambda 4534,42$ emission, which would make it an O3 by
the criteria enumerated by Walborn et al.\ (2002). In addition, N~V $\lambda 4603, 19$ absorption is
stronger than what we would expect for an O4 star.
He~II $\lambda 4542$ has an
EW of 730 m\AA, and so $\log W'=-0.61$, right on the border between O4 and earlier types.
In addition to N~III $\lambda 4634,42$
emission, He~II $\lambda$ 4686 is weak with emission wings, and so we call this
an O3.5~III(f).  Still, N~V $\lambda 4603,19$ appears to be even a bit too strong for this late
a classification; possibly the star is composite, although here we will treat it as single.
The star was previously classified by Moffat (1983) as 
considerably later, and of lower luminosity, O5.5~V.  

{\it NGC3603-38}.  The spectrum of this star is shown in Fig.~\ref{fig:spec1}.
We measure He~I $\lambda 4471$ to have an EW of 120m\AA,
while He~II $\lambda 4542$ has an EW of 590m\AA, leading to an O4 class
($\log W'=-0.69$).   There is a little emission at N~III $\lambda \lambda 4634, 42$
and  He~II $\lambda 4686$ is weak with emission wings, and so we call this an O4~III(f).
  Previously it was called
an O3~V by Drissen et al.\ (1995).  The star was slightly blended on our slit, so it is
possible that the Drissen et al.\ (1995) type is more accurate. We do see what might
be weak He$I \lambda 4387$ and $\lambda 4009$, due presumably to the slight
blend.

{\it NGC3603-101}.  Our visual impression of the spectrum of this star places it in the
range  O6-O7~V type (Fig.~\ref{fig:spec3}).  We measure EWs of 430m\AA\ and 580m\AA\ for He~I $\lambda 4471$
and He~II $\lambda 4542$, respectively, leading to a $\log W'=-0.13$ and an O6.5 type
according to Conti (1988).  The star is a dwarf, judged from the lack of any emission at
He~II $\lambda 4686$.  Very weak emission at N~III $\lambda \lambda$ 4634, 42 may be present,
and we've indicated this by adding a ``((f))" description to its luminosity class.

{\it NGC3603-Sh53}.   Our spectrum of this star is unusually noisy, with a SNR of only 120
per 2\AA\  spectral resolution element.  Nevertheless, its classification is straight forward as it is of
mid-O type, with strong He~I $\lambda 4471$ and He~II $\lambda 4542$; the former is stronger
(Fig.~\ref{fig:spec3}).
  We classify this
as an O8.5~V, having measured EWs of 640m\AA\ and 380m\AA, respectively, leading to a $\log W'=0.23$.

{\it NGC3603-104}. The spectrum of this star is shown in Fig.~\ref{fig:spec1}.
 He~I $\lambda 4471$ is very weak, with an EW of 50m\AA,
comparable to that seen in other O3 stars (Kudritzki 1980). The EW of
He~II $\lambda 4542$ is 600m\AA, leading to a $\log W'=-1.08$.
 N~III $\lambda \lambda 4634, 42$ shows weak emission, and He~II $\lambda 4686$
is weakly in absorption with emission wings, and we call this star an O3 III(f). 

 {\it NGC3603-Sh56}. The spectrum of this star is shown in Fig.~\ref{fig:spec1}.
 This too is a very early O-type star, with He~I $\lambda 4471$
 having an EW of 75m\AA.  Thus we expect this star is of O3 type.
  He~II $\lambda$ 4542 has an EW of 750m\AA, leading to
 $\log W'=-1.00$, far more negative than the $\log W'=-0.6$ used to separate O4s from
 later type, and thus consistent with our assigned type.
 N~III $\lambda \lambda 4634,42$ is weakly in emission, and He~II $\lambda 4686$
 has a strong emission component (the line appears to be almost P Cygni), suggesting 
  an O3~III(f).  However, there is a hint of double lines in our spectra, and although
 we cannot say what the spectral type is of the companion, it may be quite early too.
 We list this spectrum as an O3 III(f)+O?
 We do not include this star when determining the distance modulus in the next
 section.  Moffat (1983) classified the star a bit later (O4) and
 of lower luminosity class.  
 
 {\it NGC3603-C}.  This star shows classic WR features of the WN6 subclass, with
 strong, broad He~II~$\lambda$ 4686 emission as well as N~III $\lambda \lambda 4634,42$ and
 N~IV $\lambda 4058$ emission (Fig.~\ref{fig:spec1}).  There is absorption superimposed on the
 Balmer/Pickering lines (H$\beta$ and H$\delta$), as well as at He~II $\lambda 4542$
 absorption.  We call this WN6+abs, consistent with previous classifications.

{\it NGC3603-117}. The spectrum of this star is shown in Fig.~\ref{fig:spec2}.
The lines in this star are quite wide.  The EW of He~I $\lambda 4471$ is
500m\AA, while that of He~II $\lambda 4542$ is 690 m\AA, but with considerable
uncertainty due to the broadness of the lines.  With $\log W'=-0.14$,
the Conti (1988) criteria would lean to an O6.5 classification.  However, the
weakness of He~I $\lambda 4387$ and general appearance of the spectra suggest
a slightly earlier type.  We call the star an O6~V, with the luminosity class reflecting
the lack of emission at N~III $\lambda 4634,42$ and at He~II $\lambda 4686$.

{\it NGC3603-Sh25}.  The spectrum is shown in Fig.~\ref{fig:spec4}.
This star has long been known to be an early-type B supergiant;
Moffat (1983) classified it as a B1.5 Iab.  We would make it just slightly earlier,
about B1, based on the relative strengths of Si IV $\lambda 4089$ and Si III $\lambda 4553$.

{\it NGC3603-Sh64}.  The spectrum of this star is shown in Fig.~\ref{fig:spec2}.
This is a very early O-type star, with He~I $\lambda 4471$ just
barely discernible on our spectrum; we measure an EW of about 30m\AA, making the
star O3 (or earlier).  He~II $\lambda 4542$ has an EW  of 630m\AA, making
$\log W'=-1.3$.  The luminosity class V, with weak N~III $\lambda \lambda 4634,42$ emission but
strong He~II $\lambda 4686$ absorption.  We classify it as O3 V((f)).  

{\it NGC3603-102}.  The spectrum of this star is shown in Fig.~\ref{fig:spec3}.
The SNR of our spectrum of this star is only 100 per 2\AA\ spectral
resolution element, but fortunately it is easily classified.  We call it an O8.5~V.  He~I $\lambda 4471$
has an EW of 950m\AA, while He~II $\lambda 4542$ has an EW of 530 m\AA, leading to
a $\log W'=0.25$, in agreement with this type.  It is a dwarf.

{\it NGC3603-Sh57}.  The spectrum of this star is shown in Fig.~\ref{fig:spec1}.
The EW of He~I $\lambda 4471$ is only 30m\AA, comparable to what
Massey et al.\ (2004, 2005) found in stars as early as O2.  He~II $\lambda 4542$ has
an EW of 540m\AA, so $\log W'=-1.3$.  Thus the absorption line spectrum of this
star suggest it is O3 or earlier. The strength of N~III and He~II $\lambda 4686$ emission makes this a giant.   We
find that N~IV $\lambda 4058$ emission is comparable to N~III $\lambda \lambda 4634, 42$
emission, precluding an O2 classification (Walborn et al.\ 2002).  For giants,
Walborn et al.\ (2002) would require this fact to result in an O3.5~III(f) classification,
but we are not comfortable classifying it this late given the weakness
of He~I $\lambda 4471$ absorption, and  so we call it an O3 III(f).
It was previously called O5 III(f) by Moffat (1983).

{\it NGC3603-108}. The visual impression of this star is that it is roughly O6~V (Fig.~\ref{fig:spec2}).  
Measurements of He~I $\lambda 4471$ (EW=320m\AA) and He~II $\lambda 4542$
(EW=800 m\AA) leads to a more precise O5.5 ($\log W'=-0.40$).  There is no
emission in N~III $\lambda \lambda 4634,42$ or He~II $\lambda$ 4686, leading to the
dwarf luminosity class.  The star was slightly blended with a neighbor on the slit.
We see very weak Mg~I $\lambda 4481$ absorption, which is probably due to this
blend.

{\it NGC3603-Sh18}. Visually this is a classic O3-O4 If star (Fig.~\ref{fig:spec1}), closely resembling the 
spectral standards HD 93129A and HDE 269698 illustrated in Walborn \& Fitzpatrick (1990)\footnote{Walborn et al.\ (2002) use HD 93129A now as an example of the
O2 If type.}. The strength of N~V $\lambda 4603, 19$ is intermediate between the
two. N~IV $\lambda 4058$ emission is weaker than N~III $\lambda \lambda 4634, 42$, which would lean
the classification  towards an O3.5 If-O4 If type by the criteria listed by Walborn et al.\ (2002).
The EW of He~I $\lambda 4471$ is only 50m\AA, though, while
 He~II $\lambda 4542$ has an EW of 530 m\AA, leading to $\log W'=-1.0$, making
 this solidly an O3 type using the criteria suggested by Conti (1988).  We  compromise
 with an O3.5 If type.  The strong N~III $\lambda \lambda 4634, 42$ and He~II $\lambda 4686$
 emission leaves no doubt to its luminosity class.  The star was previously
 classified as much later, O6 If, by Moffat (1983).  Nolan Walborn (private communication) reports having independently classified the star O4~If from 
 as-yet unpublished data, in agreement with our own type.
 
 {\it NGC3603-Sh58}.  Visually this star is roughly O7-O8~V (Fig.~\ref{fig:spec3}).  We measure nearly
 equal He~I $\lambda 4471$ (EW=640 m\AA) and He~II $\lambda 4542$ (EW=480m\AA), leading to 
 a $\log W'=+0.13$, corresponding to an O8 class.  
 He~II $\lambda 4686$ is strongly in absorption, and the star is clearly a dwarf.
 The star was somewhat blended on our extraction aperture, and we find that
 the He~I lines may be broader than the He~II lines, suggesting the resulting type
 may be a composite.   We will therefore exclude it when computing the distance
 modulus in the next section.
  Melnick et al.\ (1989) listed this star as a non-member (their No.\ 30) based
 upon their measurement of a very blue color for the star ($B-V=0.47$), quite unliked the heavily
 reddened O stars members.  However, our photometry gives a color
 similar to that of the other O stars, and 
 a spectral type that confirms membership.  
  
 {\it NGC3603-Sh24}.  Visually this star is an O6~V (Fig.~\ref{fig:spec2}).  We measure an EW of 
 300m\AA\ for He~I $\lambda 4471$, and an EW of 610 m\AA\ for He~II $\lambda 4542$.
 The resulting value $\log W'=-0.31$ is borderline between an O5.5 and an O6, and
 we retain the O6~V classification.  The strength of He~II $\lambda 4686$ absorption
 makes this a dwarf.
 
 {\it NGC3603-Sh49}.  Our visual impression of the spectral type of this star is
 that of an O8 V (Fig.~\ref{fig:spec3}).  We measure an EW of He~I $\lambda 4471$ of 570m\AA,
 and an EW of He~II $\lambda 4542$ of 560m\AA, essentially identical, leading
 us to an O7.5 V type.
 
 {\it NGC3603-Sh47}. This is another early-type O star (Fig.~\ref{fig:spec2}), with weak He~I (EW of 90m\AA).
 He~II $\lambda 4542$ has an EW of 650 m\AA, and thus $\log W'=-0.86$, making the
 star O4 or earlier by Conti (1988).   There is only very weak N~III $\lambda 4638, 42$
 emission, and He~II $\lambda 4686$ is in absorption.  The star has previously been
 called an O4~V by Moffat (1983), and we retain  this type.
 
 {\it NGC3603-Sh23}.  This is a late-type O supergiant (Fig.~\ref{fig:spec4}).  The star closely matches
 the spectra of HD 152424 and 
 HD 104565 shown by Walborn \& Fitzpatrick (1990), and we
 thus classify the star as an OC9.7 Ia, where the ``C" denotes the excessively
 strong CIII $\lambda 4650$ line.  This is in substantial agreement with the O9.5 Iab
 type found by Moffat (1983).
 
 {\it NGC3603-Sh22}.  This is clearly a very early O-type star (Fig.~\ref{fig:spec1}).  Despite a signal-to-noise
 ratio of $>$350 per spectral resolution element, we detect {\it no} He~I $\lambda 4471$.
 The EW must be $<$20m\AA.  N~IV $\lambda 4058$ and N~III $\lambda \lambda 4634,42$ emission are both weak, but unfortunately the N~IV line also coincides with some
 bad pixels. We call the star an O3 III(f), although it could be an O2 by the
 Walborn et al.\ (2002) criteria.  The giant luminosity class comes about from the
 modest emission at He~II $\lambda 4686$.  The star had been called an O5 V (f)
 by Moffat (1983).
 
 {\it NGC3603-Sh 21} .  The spectrum of this star is shown in Fig.~\ref{fig:spec3}.
 Our initial impression of the spectrum of this star is that it
 is roughly of type O6~V, with He~I a bit weaker than He~II.  We measure EWs of
 390 m\AA\ and 700 m\AA, respectively, leading to a value $\log W'=-0.25$, consistent
 with the O6 class.
 
 {\it NGC3603-Sh 19}. The spectrum of this star is shown in Fig.~\ref{fig:spec2}.
 This is another very early O star.  Our spectrum has a fairly
 low SNR (200 per 2\AA\ resolution element) and we detect {\it no} He~I.
 We call this an O3~V((f)). 

\section{Results}
\label{Sec-results}

We are now prepared to derive values for the reddening, distance, and age of the NGC~3603 spectroscopic
sample.  We begin with the reddenings.  In Table~\ref{tab:results} we include the values of $E(B-V)$ we derive,
using the intrinsic colors of Table~3 from Massey (1998a), and the observed colors from Table~\ref{tab:sample}.

We derive an average reddening $E(B-V)=1.394\pm0.012$ (standard deviation of the mean, hereafter ``s.d.m.").  
The median
is 1.39.  This is very
similar to Moffat (1983)'s finding of an average value of $E(B-V)=1.44$.  
Sung \& Bessell (2004) find a somewhat
lower value, $E(B-V)=1.25$, in the core of the cluster, but argue that it increases
to larger values at greater distances.
Moffat (1983) noted that there was very little spread in the reddening among the cluster stars---his 
sample showed a dispersion of only 0.09~mag. 
We find an even smaller dispersion, 0.06~mag. We are thus reassured that the reddening
is well determined to the O stars in NGC~3603, and agree with the 
conclusion of Moffat (1983) that there is very
little variation in reddening across this cluster.  We adopt an average value of  $E(B-V)=1.39$.

It is interesting to note that the B1 supergiant, Sher 25, has a color excess that is considerably
larger than average.  This supergiant has been compared to the progenitor of SN 1987A (see Smith 2007 and Smartt et al.\ 2002), 
and is known to show circumstellar material (Brandner et al.\ 1997).

Next, let us derive a distance modulus to the cluster by adopting an absolute magnitude for each star
based upon the spectral type and luminosity class.  We adopt the values of Conti et al.\ (1983) for the O
stars, interpolating as needed; for the B supergiant (Sher 25) we adopt $M_V=-6.5$ from Humphreys \& McElroy (1984).
 If we make {\it no} correction
for the reddening to the cluster, we derive an {\it apparent} distance modulus of $19.12\pm0.09$ (s.d.m.).  The
standard deviation of this sample is 0.5~mag, which is what we expect, given that that is also the typical scatter
in $M_V$ for a given spectral type and luminosity class (Conti 1988).  The
median is 19.00.  If we exclude the giants and supergiants (as their absolute magnitudes might cover a larger
range), we determine similar values: the average is $19.21\pm0.12$ (s.d.m.), with a standard
deviation of the sample of 0.6~mag.  The median is 19.07.  We therefore adopt an {\it apparent} distance
modulus to the cluster of $19.1\pm0.1$.  

 If the extinction were normal ($R_V=3.1$) this would then correspond to a true distance modulus of 14.8,
 or 9.1~kpc.  However, Pandey et al.\ (2000) have investigated the reddening towards this cluster, and conclude
 that the reddening is anomalously high within, with a ratio of total to selective extinction $R_V=4.3$, a value that is more typical of dense environments (see, for 
 example,  Whittet 2003).
 They correct
 for reddening assuming an $E(B-V)=1.1$ for foreground (with $R_V=3.1$), and $R_V=4.3$ for the color excess
 above this value; i.e., $A_V=3.41-4.3[{\rm E(B-V)}-1.1]$.  If we were to make this correction, then the true distance
 modulus we would obtain would be 14.4, or 7.6~kpc.  
 
 The physical distance to NGC 3603 is poorly determined.  There are three methods that have been employed: 
 main-sequence fitting, spectroscopic parallaxes, and kinematic distance determinations based upon rotation models
 of the Milky Way.  We summarize the results in Tables~\ref{tab:distances} and \ref{tab:kinematic}.   Main-sequence fitting (Melnick et al.\ 1989,
 Pandey et al.\ 2000), Sung \& Bessell 2004) is probably the least certain of these for such a young cluster, given that the theoretical
 zero-age main-sequence is nearly vertical in the color-magnitude diagram.  This is particularly true in the $V-I$
 plane used by Sung \& Bessell (2004); see their Fig.~9. This problem was recognized by van den Bergh (1978), who
 obtained {\it UBV} photometry of cluster members, but noted that such data was not sufficient to determine a
 reliable distance.  (He did, however, determine a kinematic distance, as discussed below.)  See also the discussion
 in Sagar et al.\ (2001).
 The spectroscopic parallaxes (Moffat 1983, Crowther \& Dessart 1998,
 Sagar et al.\ 2001) should be more reliable.  Our sample of stars is considerably larger  than those previously 
 employed.  Of course, a key issue in deriving the physical distance is how reddening is treated.  In Table~\ref{tab:distances}
 we include the {\it apparent} distance moduli as well as the derived physical distances.  We can see that most modern
 studies derive a similar value for the apparent distance modulus (i.e., 18.6-19.1), despite the extremely large 
 range in physical distances derived (6.3~kpc to 10.1~kpc).

We can compare our distance to the kinematic distances (Table~\ref{tab:kinematic}).  Goss \& Radhakrishnan (1969) find a distance of 8.4~kpc,
but that was based upon the old assumption that the distance from the Galactic center to the sun is 10~kpc.  Correcting
to the more modern value $R_0=8.5$~kpc would result in a 7.1~kpc distance.  Van den Bergh (1978) also derives
a similar kinematic distance, 6.8~kpc, when corrected to $R_0=8.5$~kpc.  De Pree et al.\ (1999) quote a value of
$6.1\pm0.6$~kpc, but one of the co-authors, Miller Goss (private communication), writes that this value should be
revised to 7.0~kpc.  Russeil (2003)  finds a slightly larger kinematic distance,
 7.9~kpc.  While these span a range, we can conclude that
our apparent distance modulus can be reconciled with the kinematic distance moduli in the literature only if
the extinction is anomalous within the cluster, as Pandey et al.\ (2000) argue\footnote{Smartt et al.\ (2002) reject
$R_V=4.3$ for Sher 25 based on the argument that applying it leads to too high a luminosity for this star, and instead
adopt $R_V=3.7$ on somewhat arbitrary grounds.  However, they had (mis)applied $R_V=4.3$ to the entire
line of sight reddening ($E(B-V)=1.60$), rather than applying $R_V=3.1$ to the foreground reddening ($E(B-V)=1.1$)
and $R_V=4.3$ to the remainder ($\Delta E(B-V)=1.60-1.10$).  Thus, using the Pandey et al.\ (2000) reddening
leads to $A_V=5.56$~mag, not 6.88~mag, and actually smaller than
the value Smartt et al.\ (2002) adopt ($A_V=3.7\times$1.60=5.92~mag).}.

For constructing the H-R diagram we have chosen to adopt our {\it apparent} distance modulus, and then correct
individual stars for differences in the average reddening by $4.3\times \Delta E(B-V)$, where $\Delta E(B-V)$ is
the difference between the average color excess 1.39 and that for an individual star.  Since the variation in the
reddening is small, using 3.1 instead would lead to at most a difference of 0.25~mag in the final value.  
We include the absolute visual magnitudes obtained by this method
in Table~\ref{tab:results}.  In most cases the agreement between the $M_V$ (based on the spectral type
and luminosity class) and that computed from the average apparent distance modulus (corrected for
reddening above or below the cluster average) is good, within the 0.5~mag scatter we expect (Conti 1988).

\begin{figure}
\epsscale{1.0}
\plotone{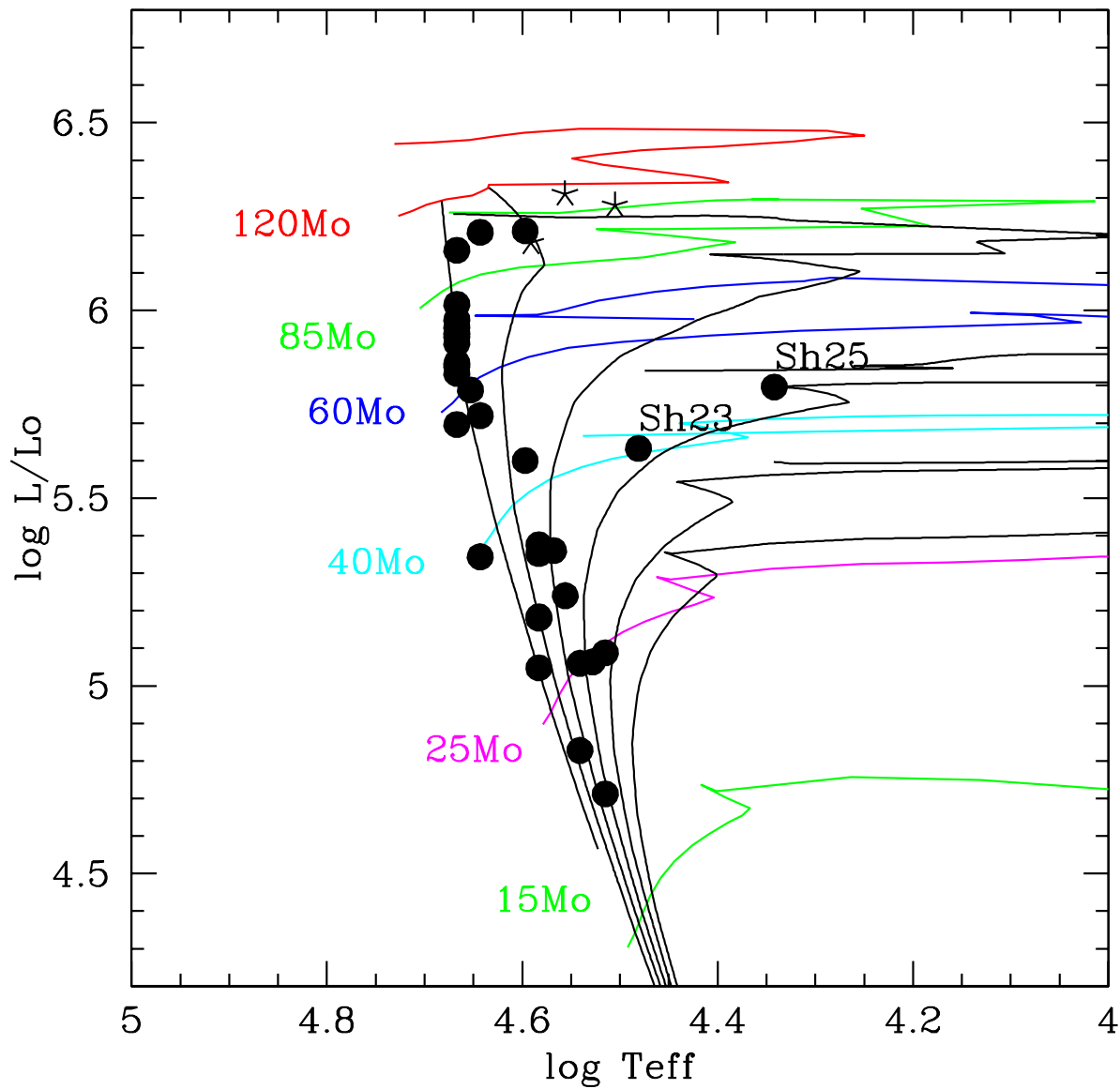}
\caption{\label{fig:hrd} The H-R diagram for NGC 3603.  Solid points show the location of the
stars listed in Table~\ref{tab:results}. The non-rotating evolutionary $z=0.020$
tracks of Schaller et al.\ (1992) are
shown, with the initial masses labeled, in various colors.  Isochrones are shown as black curves
for 1, 2, 3, 4, 5, and 6~Myr. We have labeled the locations of OC9.7 Ia star Sh23, and
the B1 Iab star Sh 25.  The three asterisks denote the location of the three H-rich Wolf-Rayet stars,
with the physical properties taken from Crowther \& Dessart (1998). }
\end{figure}

Finally, let us use the recent effective temperature scale of Massey et al.\ (2005) to place these stars on the
H-R diagram.  The new scale includes the effects of line- and wind-blanketing, which significantly lowers the
effective temperatures for Galactic stars.  For the B1 I star (Sh 25) we adopted an effective temperature of 22000~K,
consistent with the modeling of Smartt et al.\ (2002) and the recent findings of Crowther et al.\ (2006).  For that star
we have adopted a bolometric correction of -2.0~mag, also consistent with these studies.
We have included the three WR stars based upon the modeling of Crowther \& Dessart (1998), where we have
increased the luminosity by 0.12~dex for consistency with the larger apparent distance modulus found here.
We show the resulting H-R diagram in Fig.~\ref{fig:hrd}.  We compare these to
the  Schaller et al.\ (1992)
evolutionary tracks, which 
correspond to ``Galactic" (solar) metallicity ($z=0.020$), which
is appropriate for NGC 3603 
(Smartt et al.\ 2001; Peimbert et al.\ 2007; 
Lebouteiller et al.\ 2007). 
The black solid lines denote the
isochrones at 1~Myr intervals from an age of 1~Myr to 
6~Myr\footnote{We have used the older Geneva evolutionary tracks of Schaller et al.\ (1992),
 rather than the newer ones of Meynet \& Maeder (2003), simply for convenience, as
 Georges Meynet  had kindly made available software to compute
isochrones from the older tracks.  The primary difference with the newer tracks is the
inclusion of rotation.  While rotation significantly alters the tracks of low-metallicity stars,
such as those in the SMC and the LMC, there is much less of an effect at Galactic metallicities; see
Meynet \& Maeder (2000). We have also truncated the tracks at the start of the WR stage just for
clarity.}.

First, we see that the masses range above $120 M_\odot$.  We see that the most massive stars
are the stars with Wolf-Rayet features analyzed by Crowther \& Dessart (1998). 
In this context it is worth remembering that in the R136 cluster there
were several H-rich WN6 stars which had absorption lines; Massey \& Hunter (1998) concluded
that these were not evolved objects, but rather ``Of stars on steroids", i.e., stars whose masses were
so high, and which were so luminous, that their winds were so strong that their spectra 
simply {\it resembled} WR stars in having strong emission lines.  The modeling of Crowther \& Dessart (1998)
bear this out: the WRs are coeval with the rest of the cluster, and are simply slightly more luminous and
massive.  (Note that for simplicity we have truncated the evolutionary tracks prior to the Wolf-Rayet stage.)
The most massive non-WR
stars in R136 are more massive than we see here, but as discussed below, that could
be a metallicity effect.  In any event, the unevolved
stellar population of NGC~3603 is certainly quite massive.

Secondly, the ages of the highest mass stars
are roughly 1-2~Myr, with very little dispersion.  Sung \& Bessell (2004) obtained a 
similar result.
Stars of somewhat lower mass ($<40M_\odot$) show a larger dispersion in age, with
ages ranging from 1-4~Myr. The evolved B-type supergiant Sh~25 and the OC9.7 star
Sh 23 have an age of about 4 Myr.
 This is similar to the age found for stars in the outer regions of the
cluster by Sung \& Bessell (2004).

It is premature to attempt to derive an initial mass function for the high mass stars of NGC 3603,
as photometry alone cannot provide a sufficiently accurate discriminant of bolometric 
luminosity (and hence mass) at such high effective temperatures. (See discussion in Massey 1998a, 1998b.)
Sagar et al.\ 2001 and Sung \& Bessell 2004 do use their
deep photometry to compute IMFs for the intermediate-mass stars, but for these photometry does
provide a good discriminant.   In Fig.~\ref{fig:cmd}
we show the color magnitude diagram of the cluster, where we have indicated the stars for
which there are spectral types by open circles. We have cut the diagram at $V=15.6$, corresponding to a 
ZAMS 20$M_\odot$ star. (Such a star would have $M_V=3.5$ and would correspond to a late O-type
near the ZAMS.)  The photometry for the general field (shown in black) comes from Melnick et al.\ (1989), while
that of the central field (shown in red) comes from the HRC image.  We see that although we have made
significant progress in the spectroscopy of stars in NGC 3603, much work remains to be done.  In the
area outside the core, there are spectra for about 20\% of the interesting stars($V<15.6$ with colors indicating
likely membership), while in the central portion we have spectra for about a third of the stars with $V<15.6$.
(Note that all of the stars in the core have colors indicative of membership).  In all, spectra of another 90 stars
would be needed for a complete census down to 20$M_\odot$.  In the central region, observing the majority
of the remaining stars will be difficult due to crowding, but not impossible.  We hope to undertake such work
during the next observing season.

\section{Discussion and Summary}
\label{Sec-summary}

It is worth comparing what we know now about the massive star content of 
NGC~3603 with that known about the R136 cluster  
at the heart of the giant H~II region 30 Doradus in the LMC.  Spectroscopy
of the stars in R136 is similarly complete (Massey \& Hunter 1998).   R136 contains
an even greater wealth of O3 stars, and stars extending up to $150M_\odot$, about the
point where the IMF peters out to a single star. In the H-R diagram
shown by Massey
\& Hunter (1998), most of the massive stars appear to be strictly coeval, with an
age between 1 and 2~Myr, just as we find here for NGC~3603.  For stars with
masses below about 40$M_\odot$ the placement of stars in the H-R diagram of
R136 
was constrained only by photometry, and so there is an (apparent) spread in ages
from 0-6~Myr, although the actual degree of coevality may be higher.  There is a single
B-type supergiant in their H-R diagram with a mass of about 20$M_\odot$, and an
age of 10~Myr.  Maybe that star is an interloper from the field of the LMC, or it could
be that in both NGC 3603 and in R136 we simply see that a couple of high mass
stars formed a bit earlier than the majority of stars in the cluster.  

We note that both of these clusters contain H-rich WN+abs stars.  Massey \& Hunter
(1998) argued that these were {\it unevolved} (i.e., core H-burning) stars of high luminosity and mass,
whose spectra mimicked that of WN stars given the strong stellar winds expected
from high luminosity.  Given the ages of the high mass stars in NGC 3603 or
in R136, it is not possible for the WR stars to be evolved (core He-burning)
objects, not if they
are coeval with the rest of the massive stars. The physical properties of these stars
determined by modeling by Crowther \& Dessart (1998) is consistent with this\footnote{These stars were found to be {\it chemically} evolved, in the sense of showing 
enhanced CNO products at the surface, but this is still consistent with these being
H-burning objects.}.
At Galactic metallicity we would expect the onset of such
features to happen at lower mass than in the LMC, so these ``Of 
stars on steroids" in NGC 3603 may be less massive than in the R136 cluster.  

\begin{figure}
\epsscale{1.0}
\plotone{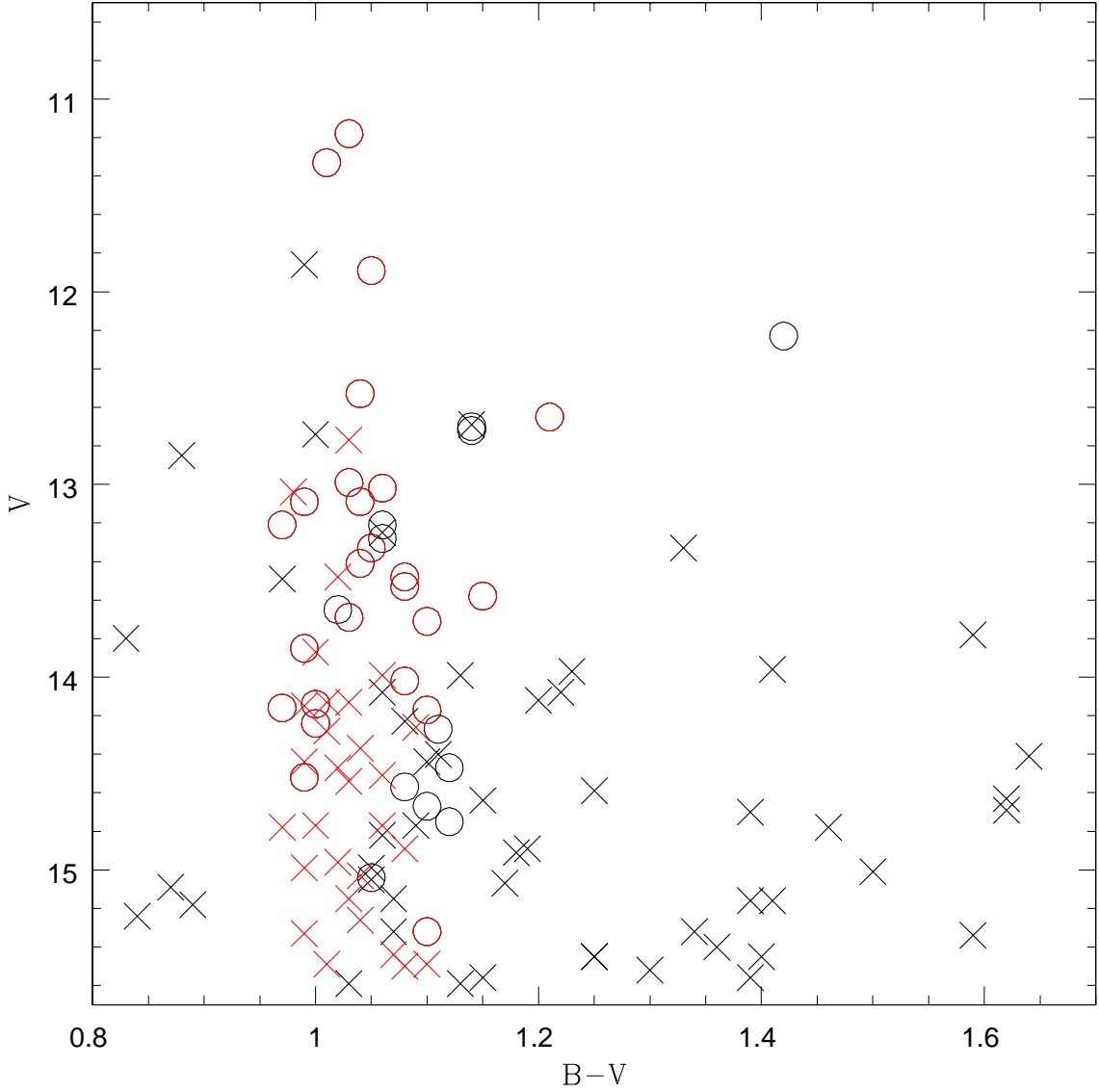}
\caption{\label{fig:cmd} The color-magnitude diagram for NGC 3603. Open circles denote the stars
for which there are spectral types; crosses denote the stars for which there are only photometry.
Red indicates the stars in the central region (i.e.,that covered  in Fig.~\ref{fig:fc2}),
and are based upon point spread function fitting to stars on the HRC images.  (This photometry
will be fully presented in a later paper.)
Black indicates photometry of stars outside the central region, from Melnick et al.\ (1989).
 }
\end{figure}

We can compare the stellar content straightforwardly. Let us simply count the
number of stars with known spectroscopy brighter than a certain bolometric
luminosity, $M_{\rm bol}\sim -10$. (For these, our spectroscopy is mostly
complete; ($V\sim$13-13.5; see Fig.~\ref{fig:cmd}). From Table~\ref{tab:results} we find 9
stars listed in NGC~3603---12, if we then
include the WN+abs stars in the tally.  There are perhaps another 6 stars that have not been
observed spectroscopically that could be as bolometrically luminous, so the total is 12-18.
In R136 we find 20-29 such stars, depending
upon the adopted temperature scale (i.e, Table~3 of Massey \& Hunter 1998).  So,
while R136 is richer in massive stars than NGC~3603, it is only by a factor of 1.1 to 2.4.
Moffat et al.\ (1994) have also argued for the similarity of the central clusters, suggesting
that the primary difference is that NGC 3603 lacks a surrounding massive halo of
cluster stars.

In summary, we have obtained spectra of 26 stars in the NGC~3603 cluster,
16 of which have no previous spectroscopy. That brings the total number of stars
with spectral types to 38 (Table~\ref{tab:sample}).  In addition, we provide
identification and
ACS/HRC photometry of another 12 stars in the central core for which spectroscopy
would be desirable.   Our spectroscopic sample includes many stars of type O3.
We find an average reddening $E(B-V)=1.39$, with very little scatter (sample
standard deviation of 0.05~mag), indicative of very little variation in reddening
in the core.  Our spectroscopic parallax for the cluster can be reconciled with the
kinematic distance of the cluster if we adopt the reddening proposed by Pandey et al.\ (2000);
we then derive a distance of 7.6~kpc.  We emphasize that although there has been
a very large range of physical distances derived for this cluster in the past 10 years
(6.3-10.1~kpc), there is excellent agreement in the {\it apparent} distance modulus
of the cluster (18.6-19.1~mag).  The disagreement in the distances are really based
upon exactly how to correct for reddening.  We side step the issue by using the
apparent distance modulus (plus a modest correction for differential reddening)
to construct the H-R diagram.  It  
reveals a mostly coeval population of massive stars extending
beyond 120$M_\odot$, with ages of 1-2~Myr.  The  most massive and luminous stars
are the H-rich WN stars, as expected by analogy with R136 (Massey \& Hunter 1998).
Some stars of 20-40$M_\odot$ may
show a larger age spread, up to 4~Myr, with the OC9.7 and B-type supergiants having an
age of 4~Myr.

Additional spectroscopy of stars is underway by A. F. J. Moffat (private communication),
which will further help refine our view of this interesting cluster.  While
further {\it HST} STIS spectroscopy of the most crowded members would be highly
desirable, it is worth noting how much can now be accomplished from the ground
using the best modern telescopes.  More such work is needed for deriving an initial
mass function for the massive stars in this cluster, as less than a third of the stars with
masses above 20$M_\odot$ have spectroscopy.  We hope to help improve this situation
in the coming observing season ourselves.

\acknowledgements

We are grateful to David Osip for allowing us to observe during some Magellan engineering time shortly in advance of our 
scheduled night.
Jes\'us Ma\'{\i}z Apell\'aniz
was kind enough to allow us access to one of the ACS/HRC frames prior
to it becoming public in order to prepare for the observing run; we
also gratefully acknowledge his calling our attention to an error we
made in the photometric transformations in an earlier version of the
manuscript.  Jennifer Mack provided guidance in our use of SYNPHOT.
We also benefited from correspondence with Geoff Clayton,
Miller Goss, 
Tony Moffat, Brian Skiff,
and
Nolan Walborn, as well as from comments on an early
draft by Deidre Hunter.  Suggestions by
the referee, Stephen Smartt, were useful in improving the paper.
NWM and PM work was supported
through the National Science Foundation (NSF) AST-0604569.  AMZ's work was supported through
an NSF REU grant, AST-0453611.

\clearpage

\begin{deluxetable}{l c c c c l l}
\tabletypesize{\footnotesize}
\tablewidth{0pc}
\tablenum{1}
\tablecolumns{7}
\tablecaption{\label{tab:sample} NGC 3603 Stars with New Data}
\tablehead{
\colhead{Star\tablenotemark{a}}
&\colhead{$\alpha_{\rm 2000}$}
&\colhead{$\delta_{\rm 2000}$}
&\colhead{$V$}
&\colhead{$B-V$}
&\colhead{Spectral Type}
&\colhead{Ref}
}
\startdata
 Sh27&11 15 03.921&-61 15 23.06&15.04&1.05\tablenotemark{b}    &O7.5 V      &1\\                                        
 Sh54&11 15 03.976&-61 15 35.72&14.57&1.08\tablenotemark{b}     &O6 V      &1\\                                         
  103\tablenotemark{c}&11 15 06.237&-61 15 36.58&13.09&0.99&O3 V((f))      &1\\                                         
  128&11 15 06.422&-61 15 36.42&14.76&0.98&\nodata   &\nodata\\                                                         
  109\tablenotemark{c}&11 15 06.588&-61 15 40.40&13.85&0.99&O7 V      &1\\                                              
 Sh50&11 15 06.615&-61 15 23.81&14.70&1.14&\nodata   &\nodata\\                                                         
  129&11 15 06.685&-61 15 38.85&14.75&0.99&\nodata   &\nodata\\                                                         
  137&11 15 06.726&-61 15 39.26&14.96&0.99&\nodata   &\nodata\\                                                         
  111&11 15 06.741&-61 15 35.64&13.87&1.00&\nodata   &\nodata\\                                                         
  124&11 15 06.815&-61 15 33.95&14.54&1.02&\nodata   &\nodata\\                                                         
 Sh63&11 15 06.847&-61 15 44.69&13.41&1.04&O3.5 III(f)    &1\\                                                          
   38\tablenotemark{c}&11 15 06.917&-61 15 36.60&13.21&0.97&O4 III(f)    &1\\                                           
   45&11 15 06.944&-61 15 38.30&14.14&1.00&O8 V-III  &2\\                                                               
   37&11 15 07.003&-61 15 37.45&14.16&0.97&O6.5+OB?  &2\\                                                               
   41&11 15 07.046&-61 15 38.95&14.24&1.00&O4 V      &2\\                                                               
   42&11 15 07.060&-61 15 39.30&12.99&1.03&O3 III(f*)&2\\                                                               
  101&11 15 07.068&-61 15 45.32&14.02&1.08&O6.5 V((f))    &1\\                                                          
   36&11 15 07.078&-61 15 37.56&14.52&0.99&O6 V      &2\\                                                               
   40&11 15 07.124&-61 15 39.09&13.33&1.05&O3 V      &2\\                                                               
 Sh53&11 15 07.148&-61 15 54.86&14.47&1.12\tablenotemark{b}     &O8.5 V      &1\\                                       
  120&11 15 07.211&-61 15 41.50&14.35&1.03&\nodata   &\nodata\\                                                         
  104&11 15 07.279&-61 15 34.94&13.02&1.06&O3 III(f)    &1\\                                                            
   A1&11 15 07.305&-61 15 38.43&11.18&1.03&WN6+abs   &2\\                                                               
   A2&11 15 07.313&-61 15 38.79&12.53&1.04&O3 V      &2\\                                                               
   A3&11 15 07.352&-61 15 38.46&13.09&1.04&O3 III(f*)&2\\                                                               
   33\tablenotemark{d}&11 15 07.363&-61 15 39.54&13.69&1.03&O5 V+OB?  &2\\                                              
    B&11 15 07.411&-61 15 38.58&11.33&1.01&WN6+abs   &2\\                                                               
 Sh56&11 15 07.498&-61 15 46.35&13.48&1.08&O3 III(f)+O?    &1\\                                                         
    C&11 15 07.589&-61 15 38.00&11.89&1.05&WN6+abs   &1\\                                                               
  139&11 15 07.591&-61 15 32.64&14.88&1.08&\nodata   &\nodata\\                                                         
  116&11 15 07.593&-61 15 35.96&14.10&1.04&\nodata   &\nodata\\                                                         
  117&11 15 07.623&-61 15 30.24&14.17&1.10&O6 V      &1\\                                                               
  125&11 15 07.630&-61 15 31.35&14.50&1.06&\nodata   &\nodata\\                                                         
 Sh25&11 15 07.649&-61 15 17.59&12.23&1.42\tablenotemark{b}     &B1 Iab    &1\\                                         
  141&11 15 07.692&-61 15 34.65&15.02&1.04&\nodata   &\nodata\\                                                         
 Sh64&11 15 07.822&-61 15 27.93&13.58&1.15&O3 V((f))      &1\\                                                          
   16&11 15 07.825&-61 15 37.84&13.53&1.08&O3 V     &2\\                                                                
  102&11 15 08.073&-61 15 34.24&15.32&1.10&O8.5 V      &1\\                                                             
  135&11 15 08.106&-61 15 40.30&14.75&1.06&\nodata   &\nodata\\                                                         
 Sh57&11 15 08.198&-61 15 47.33&13.28&1.06\tablenotemark{b}     &O3 III(f)    &1\\                                      
  108\tablenotemark{c}&11 15 08.513&-61 15 38.30&13.71&1.10&O5.5 V      &1\\                                            
 Sh18&11 15 08.712&-61 15 59.95&12.65&1.21\tablenotemark{b}     &O3.5 If     &1\\                                       
 Sh58\tablenotemark{c}&11 15 08.699&-61 15 44.49&14.24&\nodata  &O8 V\tablenotemark{e} &1\\                          
 Sh24&11 15 08.905&-61 15 27.32&14.27&1.11\tablenotemark{b}     &O6 V     &1\\                                          
 Sh49&11 15 09.129&-61 15 33.18&14.67&1.10\tablenotemark{b}     &O7.5 V &1\\                                            
 Sh47&11 15 09.353&-61 16 02.07&12.72&1.14\tablenotemark{b}     &O4 V      &1\\                                         
 Sh23&11 15 09.849&-61 15 30.48&12.70&1.14\tablenotemark{b}     &OC9.7 Ia  &1\\                                          
 Sh22&11 15 10.071&-61 15 38.01&13.21&1.06\tablenotemark{b}     &O3 III(f)    &1\\                                      
 Sh21&11 15 11.074&-61 15 36.85&14.75&1.12\tablenotemark{b}     &O6 V((f))      &1\\                                    
 Sh19&11 15 11.317&-61 15 55.63&13.65&1.02\tablenotemark{b}     &O3 V((f))      &1\\                                    
 \enddata
\tablecomments{Units of right ascension are hours, minutes, and seconds, and units
of declination are degrees, arcminutes, and arcseconds.}
\tablenotetext{a}{Nomenclature: stars with ``Sh" designations are from
Sher 1965. Stars with two digits are from Moffat et al.\ 1994, and are re-identified in Drissen et al.\ 1995 as ``HST-nn".  The ``A", ``B", and ``C" components in the central core are from Drissen et al.\ 1995.  The three-digit numbers  are from the present paper.}
\tablenotetext{b}{$B-V$ values from Melnick et al.\ 1989.}
\tablenotetext{c}{Slightly blended on slit.}
\tablenotetext{d}{Blended on ACS image.}
\tablenotetext{e}{Composite?}
\tablerefs{For spectral types: (1) This paper; (2) Drissen et al.\ 1995.}
\end{deluxetable}

\begin{deluxetable}{l l l l}
\tabletypesize{\footnotesize}
\tablewidth{0pc}
\tablenum{2}
\tablecolumns{4}
\tablecaption{\label{tab:spec}NGC 3603 Stars with Newly Found Spectral Types}
\tablehead{
\colhead{Star}
&\colhead{New }
&\multicolumn{2}{c}{Literature} \\  \cline{3-4}
& \colhead{Spectral Type}
& \colhead{Spectral Type}
&\colhead{Ref}  \\
}
\startdata
Sh27	&		O7.5 V	&\nodata		&\nodata		\\
Sh54	&		O6 V	& \nodata		&\nodata		\\
103\tablenotemark{a}	         &		O3 V((f))	&\nodata		&\nodata		\\
109\tablenotemark{a}	         &		O7 V	&\nodata		&\nodata		\\
Sh63	&       	O3.5 III(f)	 &  O5.5 V	&	1	\\
38\tablenotemark{a}	          &		O4 III(f)	& O3 V	&	2	\\
101	         &	          O6.5 V((f))	     &\nodata		&\nodata		\\
Sh53	&		O8.5 V	&	\nodata	&	\nodata	\\
104	         &		O3 III(f)	&	\nodata	&	\nodata 	\\
Sh56	&	 	O3 III(f)+O	?&	O4 V(f)	&1		\\
C	&	WN6 + abs	&	WN6 + abs	&	2\\
117	&	O6 V	&	\nodata	&	\nodata	\\
Sh25	&		B1 Iab	&	B1.5 Iab	&	1	\\
Sh64	&	O3 V((f))	&\nodata		&\nodata	\\
102	&		O8.5 V	&\nodata		&\nodata	\\
Sh57	& 	O3 III(f)	&	O5 III(f)	&	1	\\
108\tablenotemark{a}	&	O5.5 V	&\nodata		&\nodata		\\
Sh18	&  O3.5 If	&	O6 If	&	1\\
Sh58	&	O8 V	\tablenotemark{a}&\nodata		&\nodata	\\
Sh24	&		O6 V	\nodata	&\nodata	\\
Sh49	&		O7.5 V	&	\nodata	&\nodata	\\
Sh47	&	O4 V	&	O4 V	&	1	\\
Sh23	&  OC9.7 Ia	&	O9.5 Iab	&	1\\
Sh22	&	O3 III(f)	&	O5 V(f)	&	1  \\
Sh21	&	O6 V((f))	&\nodata		&\nodata		\\
Sh19	&	O3 V((f))	&	\nodata	&\nodata		\\
\enddata
\tablerefs{
References for spectral types from the literature: (1) Moffat  1983; (2) Drissen et al.\  1995.}
\tablenotetext{a}{Composite?}
\end{deluxetable}

\begin{deluxetable}{l r r l r r r r r  r}
\tabletypesize{\footnotesize
}
\tablewidth{0pc}
\tablenum{3}
\tablecolumns{10}
\tablecaption{\label{tab:results}Reddening, Distances, Absolute Magnitudes}
\tablehead{
\colhead{Star}
&\colhead{$V$}
&\colhead{$B-V$}
&\colhead{Spectral}
&\colhead{$E(B-V)$}
&\colhead{Adopted}
&\colhead{$V-M_V$}
&\colhead{Computed} 
&\colhead{$\log T_{\rm eff}$}
&\colhead{$M_{\rm bol}$} \\
&&&\colhead{Type}&&\colhead{$M_V$\tablenotemark{a}} & &\colhead{$M_V$\tablenotemark{b}}
}
\startdata
 Sh27 & 15.04 & 1.05    &O7.5 V       & 1.37      & -5.0 & 20.04 & -4.0 & 4.541 &  -7.3 \\          
 Sh54 & 14.57 & 1.08    &O6 V         & 1.40      & -5.1 & 19.67 & -4.6 & 4.583 &  -8.2 \\          
  103 & 13.09 & 0.99    &O3 V((f))    & 1.31      & -5.4 & 18.49 & -5.7 & 4.667 &  -9.9 \\          
  109 & 13.85 & 0.99    &O7 V         & 1.31      & -4.9 & 18.75 & -4.9 & 4.556 &  -8.4 \\    
      Sh63 & 13.41 & 1.04    &O3.5 III(f)  & 1.37      & -5.9 & 19.31 & -5.6 & 4.653 &  -9.7 \\          
   38 & 13.21 & 0.97    &O4 III(f)    & 1.30      & -6.0 & 19.21 & -5.5 & 4.643 &  -9.5 \\    
      45 & 14.14 & 1.00    &O8 V-III     & 1.32      & -4.8 & 18.94 & -4.7 & 4.528 &  -7.9 \\                
   41 & 14.24 & 1.00    &O4 V         & 1.32      & -5.8 & 20.04 & -4.6 & 4.643 &  -8.6 \\          
   42 & 12.99 & 1.03    &O3 III(f*)   & 1.36      & -5.8 & 18.79 & -6.0 & 4.667 & -10.2 \\          
 101 & 14.02 & 1.08    &O6.5 V((f))  & 1.40      & -5.1 & 19.12 & -5.1 & 4.568 &  -8.7 \\          
   36 & 14.52 & 0.99    &O6 V         & 1.31      & -5.1 & 19.62 & -4.2 & 4.583 &  -7.9 \\   
      40 & 13.33 & 1.05    &O3 V         & 1.37      & -5.4 & 18.73 & -5.7 & 4.667 &  -9.9 \\                 
 Sh53 & 14.47 & 1.12    &O8.5 V       & 1.43      & -4.9 & 19.37 & -4.8 & 4.515 &  -8.0 \\    
        104 & 13.02 & 1.06    &O3 III(f)    & 1.39      & -5.8 & 18.82 & -6.1 & 4.667 & -10.3 \\          
   A2 & 12.53 & 1.04    &O3 V         & 1.36      & -5.4 & 17.93 & -6.4 & 4.667 & -10.7 \\          
   A3 & 13.09 & 1.04    &O3 III(f*)   & 1.37      & -5.8 & 18.89 & -5.9 & 4.667 & -10.1 \\          
  117 & 14.17 & 1.10    &O6 V         & 1.42      & -5.1 & 19.27 & -5.1 & 4.583 &  -8.7 \\          
 Sh25 & 12.23 & 1.42    &B1 Iab       & 1.60      & -6.5 & 18.73 & -7.8 & 4.342 &  -9.7 \\          
 Sh64 & 13.58 & 1.15    &O3 V((f))    & 1.47      & -5.4 & 18.98 & -5.9 & 4.667 & -10.1 \\          
   16 & 13.53 & 1.08    &O3 V         & 1.40      & -5.4 & 18.93 & -5.6 & 4.667 &  -9.8 \\          
  102 & 15.32 & 1.10    &O8.5 V       & 1.41      & -4.9 & 20.22 & -3.9 & 4.515 &  -7.0 \\          
 Sh57 & 13.28 & 1.06    &O3 III(f)    & 1.39      & -5.8 & 19.08 & -5.8 & 4.667 & -10.0 \\          
  108 & 13.71 & 1.10    &O5.5 V       & 1.42      & -5.2 & 18.91 & -5.5 & 4.597 &  -9.2 \\          
 Sh18 & 12.65 & 1.21    &O3.5 If      & 1.53      & -6.3 & 18.95 & -7.1 & 4.597 & -10.8 \\          
 Sh24 & 14.27 & 1.11    &O6 V         & 1.43      & -5.1 & 19.37 & -5.0 & 4.583 &  -8.6 \\          
 Sh49 & 14.67 & 1.10    &O7.5 V       & 1.42      & -5.0 & 19.67 & -4.6 & 4.541 &  -7.9 \\          
 Sh47 & 12.72 & 1.14    &O4 V         & 1.47      & -5.8 & 18.52 & -6.7 & 4.643 & -10.8 \\          
 Sh23 & 12.70 & 1.14    &OC9.7 Ia     & 1.39      & -6.0 & 18.70 & -6.4 & 4.481 &  -9.3 \\          
 Sh22 & 13.21 & 1.06    &O3 III(f)    & 1.39      & -5.8 & 19.01 & -5.9 & 4.667 & -10.1 \\          
 Sh21 & 14.75 & 1.12    &O6 V((f))    & 1.44      & -5.1 & 19.85 & -4.6 & 4.583 &  -8.2 \\          
 Sh19 & 13.65 & 1.02    &O3 V((f))    & 1.35      & -5.4 & 19.05 & -5.3 & 4.667 &  -9.5 \\          
\enddata
 \tablenotetext{a}{From spectral type and luminosity class.}
\tablenotetext{b}{Computed using our apparent spectroscopic distance modulus corrected by
$4.3\times[E(B-V)-1.39]$.}
\end{deluxetable}

\begin{deluxetable}{l l c c c l}
\tabletypesize{\footnotesize
}
\tablewidth{0pc}
\tablenum{4}
\tablecolumns{6}
\tablecaption{\label{tab:distances} Stellar Distances to NGC 3603}
\tablehead{
\colhead{Study}
&\colhead{Method}
&\colhead{$E(B-V)$}
&\colhead{Adopted $R_V$}
&\colhead{Apparent DM\tablenotemark{a}}
&\colhead{Distance} \\
&&&&\colhead{(mags)} & \colhead{(kpc)}
}
\startdata
Sher 1965  & Main-sequence fitting             &1.42  & 3.0 &  17.0   & 3.5           \\
Moffat 1974 & Main-sequence fitting            &1.32 &  3.1 &  18.7  & 8.1          \\
Melnick \& Grosbol 1982  &Main-sequence fitting & 1.38 & 3.1 & 17.9 &    5.3     \\
Moffat 1983 & Spectroscopic parallax          &1.44  & 3.2 &  18.8   &$7.0\pm0.5$ \\
Melnick et al 1989 & Main-sequence fitting  & 1.44  & 3.2 &  18.9   &7.2 \\
Crowther \& Dessart 1998 & Spectroscopic parallax & 1.23 & 3.2 & 18.8 & 10.1\\
Pandey et al.\ 2000 & Main-sequence fitting & 1.48 & 3.1/4.3\tablenotemark{b} & 19.0 & $6.3\pm0.6$\\
Sagar et al.\ 2001 & Spectroscopic parallax &  1.44 & 3.1  & 18.8 &  $7.2\pm1.2$ \\
Sung \& Bessell 2004 & Main-sequence fitting & 1.25\tablenotemark{c} & 3.55 & 18.6 & $6.9\pm0.6$ \\
This Study    & Spectroscopic parallax                                                     &1.39 & 3.1/4.3\tablenotemark{b} & 19.1 & 7.6 \\
\enddata
\tablenotetext{a}{Apparent distance modulus computed using the quoted true distance modulus and the
reddening correction made in each study.}
\tablenotetext{b}{A normal $R_V$ of 3.1 is applied for the foreground reddening $E(B-V)=1.1$, with
a value of $R_V$ applied to extinction within the cluster, i.e., $A_V=3.1 \times 1.1 + [(E(B-V)-1.1)\ \times 4.3]$}
\tablenotetext{c}{For the central cluster.}

\end{deluxetable}

\begin{deluxetable}{l l }
\tabletypesize{\footnotesize
}
\tablewidth{0pc}
\tablenum{5}
\tablecolumns{6}
\tablecaption{\label{tab:kinematic} Kinematics Distances to NGC 3603}
\tablehead{
\colhead{Study}
&\colhead{Distance (kpc)}
}
\startdata
Goss \& Radhakrishnan 1969  & 7.1\tablenotemark{a} \\
van den Bergh 1978           & $6.8\pm0.9$\tablenotemark{a}  \\
De Pree et al.\ 1999 &  7.0\tablenotemark{b} \\
Russeil 2003             &  7.9 \\ 
\enddata
\tablenotetext{a}{Corrected to $R_0=8.5$~kpc.}
\tablenotetext{b}{De Pree et al.\ 1999 quote a value of 6.1, but one of the co-authors,
Miller Goss (private communication), writes that the value was a mistake, and recomputes a value of 7.0~kpc.}
\end{deluxetable}

\vskip -20pt
\begin{references}
\reference {} Brandner, W., Grebel, E. K., Chu, Y.-H., \& Weis, K. 1997, ApJ, 475, L45
\reference {} Bessell, M. S. 1990, PASP, 102, 1181
\reference {} Bessell, M. S., Castelli, F., \& Plez, B. 1998, A\&A, 333, 231
\reference {} Buser, R., \& Kurucz, R. L. 1978, A\&A, 70, 555
\reference {} Cardelli, J. A., Clayton, G. C., \& Mathis, J. S. 1989, ApJ, 345, 245
\reference {} Conti, P. S. 1973, ApJ, 179, 181
\reference {} Conti, P. S. 1988, in O Stars and Wolf-Rayet Stars, ed.\  P. S. Conti and A. B. Underhill (Washington, DC: NASA SP-497)
\reference {} Conti, P. S., \& Alschuler, W. R. 1971, ApJ, 170, 325
\reference {} Conti, P. S., \& Frost, S. A. 1977, ApJ, 212, 728
\reference {} Conti, P. S., Garmany, C. D., de Loore, C., \& Vanbeveren, D. 1983, ApJ, 274, 302
\reference {} Crowther, P. A., \& Dessart, L. 1998, MNRAS, 296, 622
\reference {} Crowther, P. A., Lennon, D. J., Walborn, N. R. 2006, A\&A, 446, 279
\reference {} De Pree, C., G., Nysewander, M. C., \& Goss, W. M. 1999, AJ, 117, 2903
\reference {} Drissen, L., Moffat, A.F.J., Walborn, N.R., \& Shara, M.M., 1995, AJ, 110, 2235
\reference {} Goss, W. M., \& Radhakrishnan, V. 1969, Ap Letters, 4, 199
\reference {} Harayama, Y., Eisenhauer, F., \& Martins, F. 2007, ApJ, in press
\reference {} Eisenhauer, F., Quirrenbach, A., Zinnecker, H., \ Genzel, R. 1998, ApJ, 498, 278
\reference {} Humphreys, R. M., \& McElroy, D. B. 1984, ApJ, 565
\reference {} Hunter, D. A., O'Neil, E. J., Lynds, R., Shaya, E. J., Groth, E., \& Holtzman, J. A. 1996, ApJ, 459, L27
\reference {}  Kudritzki, R. P. 1980, A\&A, 85, 174
\reference {} Kurucz, R. 1992, in The Stellar Populations of Galaxies, ed.\ B. Barbuy
\& A. Renzini (Dodrecht: Kluwer), 225
\reference {} Lebouteiller, V., Bernard-Salas, J., Brandl, B., Whelan, D.,
Wu, Y., Chamandaris, V., \& Devost, D. 2007, ApJ, submitted; astro-ph/0710.4549
\reference {} Ma\'{\i}z Apell\'aniz, J. 2006, AJ, 131, 1184
\reference {} Massey, P. 1998a, in Stellar Astrophysics for the Local Group, ed.\ A. Aparicio, A.
Herrero, \& F. Sanchez (Cambrdige: Cambridge Univ.\ Press), 95
\reference {} Massey, P. 1998b, in The Stellar Initial Mass Function, 38th Herstmonceux Conference, ASP Conf.\ Ser.142,
ed. G. Gilmore \& D. Howell (San Francisco, ASP), 17
\reference {}  Massey, P., Bresolin, F., Kudritzki, R. P., Puls, J., \& Pauldrach, A. W. A. 2004, ApJ 608, 1001
\reference {} Massey, P., \& Hunter, D. A. 1998, ApJ, 493, 180
\reference {}  Massey, P., Puls, J., Pauldrach, A. W. A., Bresolin, F.,
Kudritzki, R. P., \& Simon, T. 2005, ApJ, 627, 477
ApJ 634, 1286
\reference {} Melnick, J., \& Grosbol, P. 1982, A\&A, 107, 23
\reference {} Melnick, J., Tapia, M., \& Terlevich, R. 1989, A\&A, 213, 89 
\reference {} Meynet, G., \& Maeder, A. 2000, A\&A, 361, 101
\reference {} Meynet, G., \& Maeder, A. 2003, A\&A, 404, 975
\reference {} Moffat, A. F. J. 1974, A\&A, 35, 315
\reference {} Moffat, A.F.J., 1983, A\&A, 124, 273
\reference {} Moffat, A. F. J., Drissen, L., \& Shara M. M., 1994, ApJ, 436, 183
\reference {} Pandey, A. K., Ogura, K., \& Sekiguchi, K. 2000, Publ.\ Astron.\ Soc.\ Japan 52, 847
\reference {} Pavlovsky, C. et al.\ 2006, {\it ACS Data Handbook}, Version 5.0
(Baltimore: STScI)
\reference {} Peimbert, M., Peimbert, A., Esteban, C., Garcia-Rojas, J.,
Bresolin, F., Carigi, L., Ruiz, M. T., Lopez-Sanchez, A. R. 2007, RMxAC, 29, 72
\reference {} Russeil, D. 2003, A\&A, 397, 133
\reference {} Sagar, R., Munari, U., \& de Boer, K. S. 2001, MNRAS, 327, 23
\reference {} Schaller, G., Schaerer, D., Meynet, G., \& Maeder, A. 1992, A\&AS, 96, 269
\reference {} Sher, D. 1965, MNRAS, 129, 237 
\reference {} Simon, K. P.,  Kudritzki, R. P., Jonas, G., \& Rahe, J. 1983, A\&A, 125, 34
\reference {} Sirianni, M. et al., 2005, PASP, 117, 1049
\reference {} Smith, N. 2007, AJ, 133, 1034
\reference {} Smartt, S. J., Lennon, D. J., Kudritzki, R. P., Rosales, F., Ryans, R. S. I., \& Wright, N. 2002, A\&A 391, 979
\reference {} Stolte, A., Brandner, W., Brandl, B., Zinnecker, H., \& Grebel, E. K. 2004,
AJ, 128, 765
\reference {} Sung, H., \& Bessell, M. S. 2004, AJ, 127, 1014
\reference {} van den Bergh, S. 1978, A\&A, 63, 275
\reference {} Walborn, N. R., 1973, ApJ, 182, L21
\reference {} Walborn, N. R. et al.\ 2002, AJ, 123, 2754
\reference {} Walborn, N. R., \&  Fitzpatrick, E L., 1990, PASP, 102, 379
\reference {} Whittet, D. C B. 2003, Dust in the Galactic Environment, second edition
(Brisol: IOP), 91
\reference {} Zacharias, N. et al.\ 2004, AJ, 127, 3043


\end{references}
\end{document}